\documentclass[12pt]{article}
\usepackage{amsmath}
\usepackage{times}
\usepackage{graphicx}
\usepackage{color}
\usepackage{multirow}
\usepackage{epsfig,graphics,amssymb,subfigure,verbatim,lscape,fullpage}
\usepackage[authoryear]{natbib}
\usepackage{rotating}
\usepackage{bbm}
\usepackage{latexsym}

\usepackage{etoolbox}
\makeatletter
\patchcmd{\@footnotetext}{\footnotesize}{}{}{}
\makeatother

\newcommand{\beq}{\begin{equation*}}
\newcommand{\eeq}{\end{equation*}}
\newcommand{\beqr}{\begin{eqnarray}}
\newcommand{\eeqr}{\end{eqnarray}}
\newcommand{\beqrn}{\begin{eqnarray*}}
\newcommand{\eeqrn}{\end{eqnarray*}}
\newcommand{\beqn}{\begin{equation*}}
\newcommand{\eeqn}{\end{equation*}}
\newcommand{\bei}{\begin{itemize}}
\newcommand{\beii}{\begin{itemize} \item}
\newcommand{\eei}{\end{itemize}}
\newcommand{\ben}{\begin{enumerate}}
\newcommand{\een}{\end{enumerate}}
\newcommand{\bes}{\begin{small}}
\newcommand{\ees}{\end{small}}
\newcommand{\bec}{\begin{center}}
\newcommand{\eec}{\end{center}}

\newcommand{\captionfonts}{\normalsize}

\newcommand{\phr}[1]{\textcolor{magenta}{[Eric:  can we rephrase that wording?]}}

\newtheorem{proposition}{Proposition}

\makeatletter  
\long\def\@makecaption#1#2{%
  \vskip\abovecaptionskip
  \sbox\@tempboxa{{\captionfonts #1: #2}}%
  \ifdim \wd\@tempboxa >\hsize
    {\captionfonts #1: #2\par}
  \else
    \hbox to\hsize{\hfil\box\@tempboxa\hfil}%
  \fi
  \vskip\belowcaptionskip}
\makeatother   

\begin{document}
\hspace{13.9cm}

\ \vspace{20mm}\\

\noindent {\LARGE Neutral stability, rate propagation, and  \\ critical branching in  feedforward networks}


\ \\
{\bf \large Natasha Cayco Gajic$^{\displaystyle 1}$ and Eric Shea-Brown$^{\displaystyle 1}$}\\
{$^{\displaystyle 1}$University of Washington.}\\
%


\thispagestyle{empty}
\markboth{}{NC instructions}
%

%
\begin{center} {\bf Abstract} \end{center}
Recent experimental and computational evidence suggests that several dynamical properties may characterize the operating point of functioning neural networks:  critical branching, neutral stability, and production of a wide range of firing patterns. We seek the simplest setting in which these properties emerge, clarifying their origin and relationship in random, feedforward networks of McCullochs-Pitts neurons.  Two key parameters are the thresholds at which neurons fire spikes, and the overall level of feedforward connectivity.  When neurons have low thresholds, we show that there is always a connectivity for which the properties in question all occur:  that is, these networks preserve overall firing rates from layer to layer and produce broad distributions of activity in each layer.  This fails to occur, however, when neurons have high thresholds. A key tool in explaining this difference is eigenstructure of the resulting mean-field Markov chain, as this reveals which activity modes will be preserved from layer to layer. 
We extend our analysis from purely excitatory networks to more complex models that include inhibition and ``local" noise, and find that both of these features extend the parameter ranges over which networks produce the properties of interest. 



\section{Introduction}

Many basic questions remain unresolved in understanding how simple features of network connectivity determine the statistical structure of their outputs.   In particular, as we vary the average connectivity strength between model neurons, what kinds of transitions occur in model dynamics? The first dynamical property we might study at a transition is neutral stability of trajectories. Intuitively, it appears that neutral stability could favor signal transmission, because it suggests that input patterns (and their noisy perturbations) will retain their original separation in state space, neither diverging nor converging towards some fixed attractor~\citep{Bertschinger, Legenstein}. The second, allied property that could occur as networks transition from weak to strong connectivity is the production of a wide range of output states -- that is, a mix of firing patterns that induce a broad distribution with high response entropy.  If responses are tallied via total network output, this could require statistical correlations of all orders~\citep{Amari}; thus, higher-order correlations are another statistical property of interest at network transitions. Finally, an assay that involves all of these properties is the decodability of input patterns based on network outputs.

But how are all of these properties related?  Do networks ever exhibit all of them simultaneously, and if so, when?  Developing the complete picture is a formidable challenge; in this paper, we progress by answering these questions in what is probably the most tractable class of systems in which they can be studied.  These are noisy, feedforward networks of thresholding neurons~\cite{Nowotny} .  

Several important prior studies of signal propagation in feedforward networks inform our approach. These suggest that a wide range of network responses fails to occur in broad parameter regimes:  rather, the only outputs produced are all cells firing or being silent simultaneously.  This is due to the buildup of correlations among neural activity at each layer, even when inputs drive the cells to fire independently in the first layer. In particular, for iteratively constructed \emph{in vitro} feedforward networks, neurons displayed a marked tendency to synchronize~\citep{Reyes}. Subsequent simulations and analyses with thresholding neurons have corroborated these findings, arguing that any initial spike count distribution  becomes strongly bimodal after a few layers \citep{Nowotny}. Integrate-and-fire neurons similarly fail to transmit rates without decaying or saturating to a point independent of the input rate, but are able to support propagation of highly-synchronized volleys of spikes~\citep{Litvak}. In contrast, different studies demonstrate a ``critical" regime with broadly distributed output patterns and significant higher-order interactions \citep{Beggs, Yu}. 

As we will further explore here, the key difference among these studies turns out to be the {\it threshold} number of excitatory inputs that each cell must receive in order to fire~\citep{Kumar}.  This threshold is low (a single spike) in the work of \citet{Beggs} but much higher for \citet{Nowotny}, \citet{Reyes}, and \citet{Litvak}. As reviewed in \citet{Kumar}, densely connected feedforward networks with synapses that are weak relative to threshold tend to produce more synchrony than their sparsely connected counterparts, due to the neurons having more shared inputs~\citep{Rosenbaum}. Thus, as synaptic inputs are weak compared to spike thresholds in many biological settings, it may appear that synchrony is inevitable.   However, noise ``local" to each neuron decreases synchrony, and can do so  without damaging the capacity to transmit signals, at least those defined by firing rates within each network layer (\citet{van Rossum}, but see \cite{Nowotny,Reyes}).

Here, we undertake to unify  these results through a common mathematical framework, and extend them by treating multiple assays of network outputs.  In particular, we show when and how neutral stability, broad response distributions, higher-order correlations, and the transmission of firing rate signals {coexist} and when these properties fail to coexist. For any level of spike threshold, we find that there is always an intermediate value of network connectivity characterized by neutral stability and higher order correlations.  High response entropy and transmission of firing rates, however, only occur at this point when neurons have low thresholds or added noise.  

The narrative of the paper proceeds as follows. Section 2 gives the setup of the model. In Section 3, we introduce the  branching ratio, describing how layer-averaged activity -- that is, firing rates --  are propagated from layer to layer.  Next, in Section 4, we develop tools that give a more refined view of how activity is transmitted.  Specifically, we show that the model can be reduced to a mean-field Markov chain, and that the eigenstructure of the corresponding transition matrix reveals intermediate parameter values for which the networks support persistent, broadly distributed responses.  In Section 5 we study the resulting responses in terms of higher-order correlations and response entropy, showing that they are both maximized at this intermediate parameter regime. Section 6 introduces a combined metric, which assess the capacity of networks to preserve rates from one layer to the next while maintaining broadly distributed responses.  In Sections 7 and 8, we apply the same analyses to excitatory-inhibitory networks and to those with localized ``background" noise, and see that both factors increase parameter ranges over which the propagation of broad activity distributions with preserved firing rates occurs.


\begin{figure}[h!]
\hfill
\begin{center}
\includegraphics[width=5.7in]{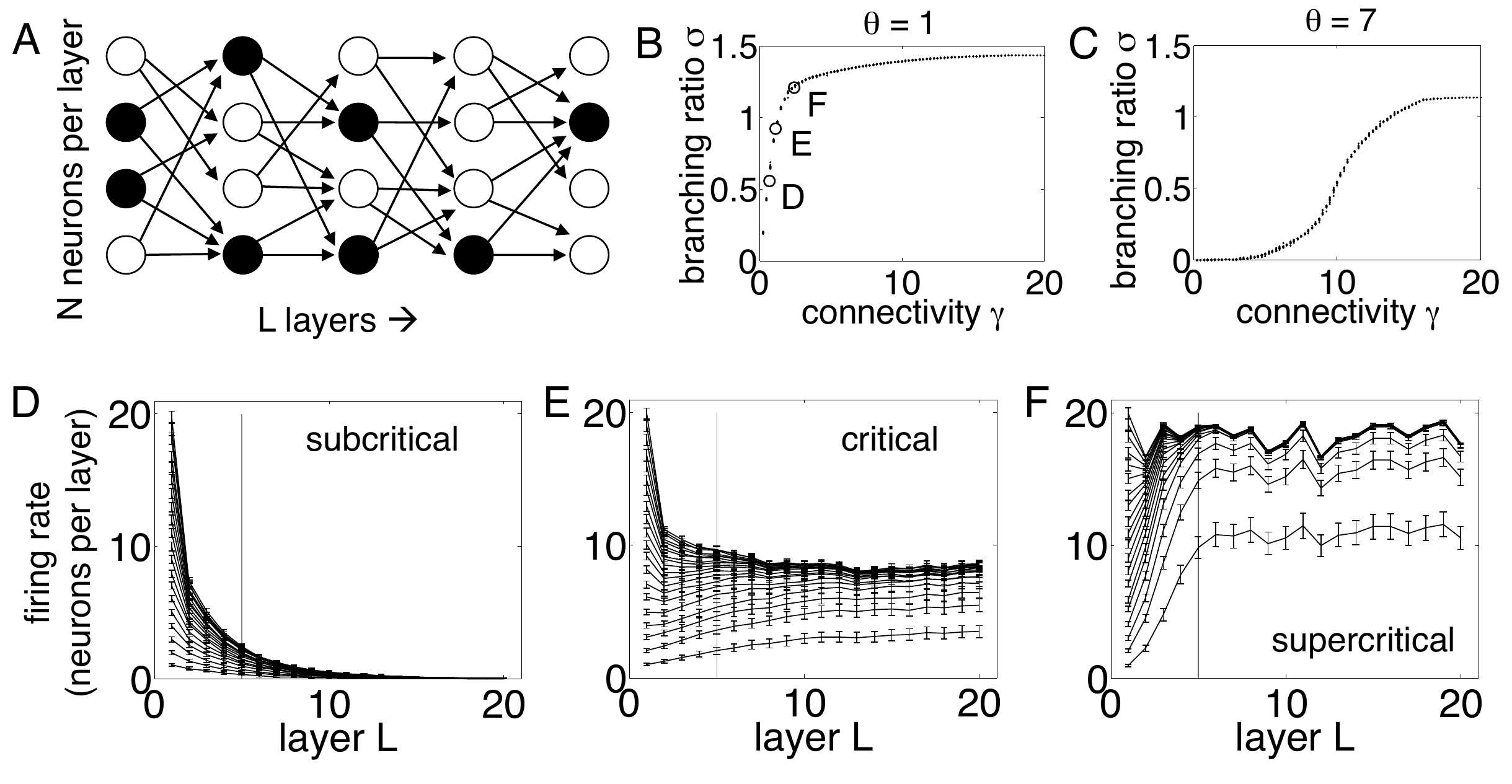}
\end{center}
\caption{Average rate transmission in feedforward networks. (A) A schematic of a feedforward network. Filled circles indicate spikes, hollow circles quiescence (i.e., absence of firing). In this example, $N = 4$, $L = 5$, $C = 2$, $p = 0.5$, and $\theta = 1$. (B, C) Branching ratio $\sigma$ as a function of connectivity strength $\gamma$ for $N = 20$, (B) $\theta = 1$ and (C) $\theta = 7$. Each data point is the branching ratio of a network of a particular connectivity structure. (D - F) Simulated propagation of firing rates shown for three sample networks with $\theta = 1$ and $C = 3$, $p = 0.25$, $0.43$, $0.85$, respectively. These parameters are also indicated by the markers in (B). Noisy, uncorrelated input is injected into the first layer, and the resulting firing rates are averaged over $1000$ trials plotted over multiple layers. Error bars indicate standard deviation scaled by a factor of 1/10 to facilitate comparison. Vertical grey bars are shown at $L = 5$ to emphasize that henceforth we will primarily be concerned with shallow layers. (D) In subcritical networks ($\sigma < 1$), activity tends to die out. (F) In supercritical ($\sigma > 1$) networks, activity saturates. (E) Critical networks ($\sigma \approx 1$) reveal greatest fidelity in propagating Poisson input rates through layers; however, while this picture is qualitatively true for networks of low-threshold neurons, when $\theta$ reaches higher values, networks tend to transmit only high or low rates (see Section 6).}
\label{schematic}
\end{figure}

\section{Model of stochastic feedforward networks}

\noindent {\it Network:}  Following \citet{Nowotny}, we examine a network of binary \citep{McCulloch} neurons in a feedforward architecture with probabilistic synapses and input (see Figure~\ref{schematic}A for a schematic). Each layer consists of $N$ identical neurons. In general, we will illustrate $N = 20$; results hold for larger $N$ as well, as we summarize in the Discussion. The neurons are thresholding units that spike if they receive at least $\theta$ synaptic inputs from neurons in the previous layer and are otherwise quiescent (i.e., silent). The connectivity structure between layers is random and spatially homogenous; each neuron upstream is connected to $C$ postsynaptic neurons chosen uniformly from the downstream layer. Connections between neurons have a fixed probability $p$ of synaptic transmission. 

\smallskip

\noindent {\it Stimuli:}  The networks are driven by a stimulus to elicit an average spike count of $S\in\{0,\dots N\}$ firing neurons in the first layer at that time step. Unless otherwise specified, this stochastic input is injected {\it independently} so that each neuron in the first layer responds as an independent (0,1) Bernoulli random variable with biased probability $S/N$ of spiking (taking value 1).  This results in a binomially distributed spike count in the first layer.

\smallskip

\noindent {\it Propagation:} The state of the $L$th layer is denoted by $\mathbf x_L$, an $N$-vector of zeros and ones, and the connectivity matrix between layers $L$ and $L+1$ by $E_L$. (Henceforth we will use $E$ to refer to the $N \times N \times L-1$ connectivity matrix of the entire network.) Since the connections between neurons are stochastic, in a given trial each synapse fails with probability $1-p$. It will be useful to call the realization of $E_L$ according to the probability of synaptic transmission the ``effective" connectivity matrix $\hat{E}_L$, keeping in mind that different trials will yield different $\hat E_L$ yet $E_L$ will remain fixed. The state at layer $L$ of a realization of a given network can now be expressed as
\begin{equation*} \mathbf x_{L+1} = \Theta(\hat E_{L} \mathbf x_L-\theta), \end{equation*}
where $\Theta$ is the elementwise Heaviside step function. The key parameter in this system is the connectivity strength $\gamma= Cp$. 

\smallskip

\noindent {\it Limitations and simplifications:} 
We note several important facts about the model setup and analysis. First, this model has no time in its dynamics; each trial can be thought of as a wave of activity evolving from a particular initialization in the first layer, and is independent of the next. Because of this, the phenomenon of synchrony in the usual temporal sense is not applicable. The corresponding concept of synchrony is when neurons in a layer tend to fire, or be quiescent, together in a given trial; this is what we will mean in the remainder of this paper when we refer to synchrony or synchronous coding. Second, because of the assumption of spatial homogeneity both in inputs and in network architecture, this model is not well-suited to study spatial modes of activity. 

Third, and most importantly, our analysis henceforth focuses on the total activity within each layer.  That is, rather than quantify network responses in the full space of $2^N$ firing patterns that can occur in each layer, we restrict our description to the {\it number} of cells that fire in that layer:  the $N+1$ different values of the (layer-averaged) firing rate.


\section{The branching ratio}

To understand the qualitative dynamics and average firing rate transmission through multiple layers, we borrow a useful tool from the criticality literature \citep{Beggs}. A ``critical" transition regime is often experimentally defined via the \emph{branching ratio} $\sigma$, the ratio between the number of cells in a population firing at a particular time step and the number of cells firing at the previous timestep, averaged over time. To avoid decay or growth of activity, the system must produce firing rate dynamics which are neutrally stable, satisfying $\sigma \approx 1$; such networks are labeled critical. 

In our feedforward framework, the relevant measurement is the branching ratio averaged over trials and layers rather than time. To quantify the general capacity of a particular network with fixed connectivity structure $E$ to maintain activity in a one-to-one manner, we will also average this layerwise branching ratio over repeated trials with the same network, each with different stimulus rates as well as different (random) inputs $\mathbf x_1$ to the first layer.  The result is:
\begin{equation*} \sigma = \left \langle \left \langle  \left \langle \frac{S_L}{S_{L-1}} \right \rangle_L \right \rangle_{\hat E, \mathbf x_1} \right \rangle_S, \end{equation*}  
where $S_L$ is the number of neurons spiking in layer $L$ on a given trial. Throughout this paper, when we refer to the branching ratio we will mean $\sigma$.

We conducted Monte Carlo simulations to compute how this quantity changes with connectivity level $\gamma$. In detail, at a fixed $\gamma$, we first chose one example of a network structure $E$ for every $C \geq \lceil \gamma \rceil$, the constrained value ensuring that $p < 1$. For each $E$, we then input a deterministic rate of exactly $S = \theta +1, \dots, N$ spiking neurons in the first layer with 100 random instantiations of $\mathbf x_1$, evolve the network, and measure the ratios ${S_L}/{S_{L-1}}$ for each layer until either the neural activity dies out or until the last layer is reached.  Finally, $\sigma$ is computed as the average over the 100 random network realizations and instantiations at the first layer, and subsequently over all stimulus levels greater than $\theta$ spiking neurons per layer (as any input less than that is guaranteed to be extinguished at the next layer.)  

Figure~\ref{schematic}BC shows results over five layers. Each of the tight scatter of points at each value of connectivity $\gamma$ is the branching ratio of a particular network with that value of connectivity and a specific architecture $E$.  (The fact that there is very little variation at a given level $\gamma$ supports our choice of this combined parameter as the principal one in our study.\footnote{In more detail: in following sections we reduce the two parameters $C$ and $p$ dictating network connectivity to the single connectivity parameter $\gamma$; this is supported by the observation that variation in $\sigma$ for fixed $\gamma$ but varying values of $C$ and $p$ is has a negligible impact on the branching ratio, as shown by the tight scatter of points at each $\gamma$ in Figure~\ref{schematic}BC.  Moreover, in the mean-field theory we develop, it is also the case that $\gamma$, rather than $C$ and $p$ separately, enters.})  Note that as we sweep connectivity $\gamma$ from small to large values, we pass through a value $\gamma_{\text{obs}}$ at which $\sigma \approx 1$.  Thus, we find that the transition (critical) branching parameter is consistently found in our networks at some intermediate connectivity level. 

We next illustrate the implications of the branching parameter for propagation of firing rates across network layers. For many different networks, we compute rate trajectories averaged over $1000$ trials for input rates ranging from 0 to $N = 20$ neurons firing in the first layer. In each trial, the input rate $S$ and $E$ are fixed, yet $\mathbf x_1$ and $\hat E$ change probabilistically. The evolution of the firing rate over $20$ layers is shown for three representative networks with threshold $\theta = 1$ in Figure~\ref{schematic}D-F. In subcritical networks (Figure~\ref{schematic}D) neural activity dies after a few layers regardless of stimulus. The supercritical, i.e. $\sigma>1$, network (Figure~\ref{schematic}E) inflates rates to nearly maximal values, and as in the subcritical case it is difficult to distinguish between two inputs based on output rate alone. In critical networks, however, rate trajectories remain separated at each layer (Figure~\ref{schematic}F). This result is in agreement with other findings in the literature regarding information transmission of critical networks.
Overall, these simulations confirm the expected picture that the average firing rate statistic is best propagated through networks when $\sigma \approx 1$. 

Beyond preservation of firing rates from one layer to the next, we are interested in networks that can produce a broad distribution of responses, and avoiding the limitations of strong synchrony.  To assess this, in the next section we introduce a tool to describe propagation of firing rate distributions across layers via a mean-field approximation.


\section{Mean-field Markov chain model and spectral analysis}

Since the state of each layer depends solely on the state of the previous layer and the synaptic connections between layers, our feedforward networks are Markov chains \citep{Nowotny}. Furthermore, as we aim to describe only the propagation of layer-averaged firing rates, rather than particular firing patterns (or binary ``words"), our Markov chain has $N+1$ states.  We proceed to derive a mean-field description of the Markov chain for each connectivity level $\gamma$ by averaging over possible connection matrices.  This mean-field model becomes exact in the special case of all-to-all connectivity ($C=1$), for which \citet{Nowotny} developed the same description.  For the excitatory networks considered in the main part of the paper (Sections 1-6), we have verified numerically that the mean-field model is a good predictor of the true spike count dynamics except in the deterministic limit of large $p$ and small $C$ (see Appendix). 
 
The mean-field transition matrix $A$ -- i.e., the matrix whose $(n,m)$th element is the probability that there are $S_L = m$ neurons spiking at a layer given $S_L = n$ spiking in the previous layer -- is given by:
\begin{equation*}
A_{nm}={N \choose m} q_n^m (1-q_n)^{N-m}
\end{equation*}
if $ n \geq \theta$.  Here, $q_n$ is the probability that a downstream neuron will fire assuming $n$ spiking neurons in the previous layer:
\begin{equation*}
q_n=\sum_{k=\theta}^n {n \choose k} \left( \frac{\gamma}{N}\right)^k \left(1-\frac{\gamma}{N}\right)^{n-k}.
\end{equation*}
If $n<\theta$, then $q_n<0$ and $A_{nm} = 0$.  The mean-field transition matrix $A$ can be derived from the transition matrix of the original Markov chain (see Appendix for details). 

Using the transition matrix, the spike count probability distribution $P_L$ at layer $L$ (the vector of length $N+1$ whose $j$th component is the probability that $j$ neurons are firing in layer $L$) is simply given by matrix-vector multiplication: $P_L = P_{L-1} A$. This averaged system is inherently permutation-symmetric due to a lack of spatial structure in the network connectivity.

\medskip

The long-term behavior of these feedforward networks can be predicted using the eigenvalues and eigenvectors of the mean-field transition matrix. To illustrate this, assume $A$ is diagonalizable, so that the input probability distribution $P_{\text{input}} = P_1$ can be decomposed into a linear combination of the eigenvectors of $A$:
\begin{equation*} P_{\text{input}} = \alpha_0 v_0+ \alpha_1 v_1 + \cdots + \alpha_{N} v_N.\end{equation*}
The spike count probability distribution at the $L$th layer is simply $P_L = P_{\text{input}} A^L$:
\begin{equation*} P_L = \alpha_0 \lambda^L v_0+ \alpha_1 \lambda_1^L v_1 + \cdots + \alpha_{N} \lambda_N^L v_N. \end{equation*}
The persistence of different eigenmodes through layers is determined by the size of their corresponding eigenvalues.\footnote{In Markov chains with very non-normal transition matrices, transient activity can persist past that expected solely by the spectrum; these matrices can be analyzed through their prominent pseudospectral sets, which are the eigenvalues of small perturbations of the matrix. When we pursued this type of analysis of $A$, we did not find significant pseudospectral sets that described the persistent activity of our networks beyond expectations from the spectral analysis (results not shown; see \citet{Trefethen} for more details on pseudospectral analysis).} If $\lambda_i \ll 1$ then after a few layers the contribution $i$th eigenmode will decay to become negligible. On the other hand, eigenmodes whose eigenvalues are near $1$ will survive through many layers.

\begin{figure}[h!]
\hfill
\begin{center}
\includegraphics[width = 5.2 in] {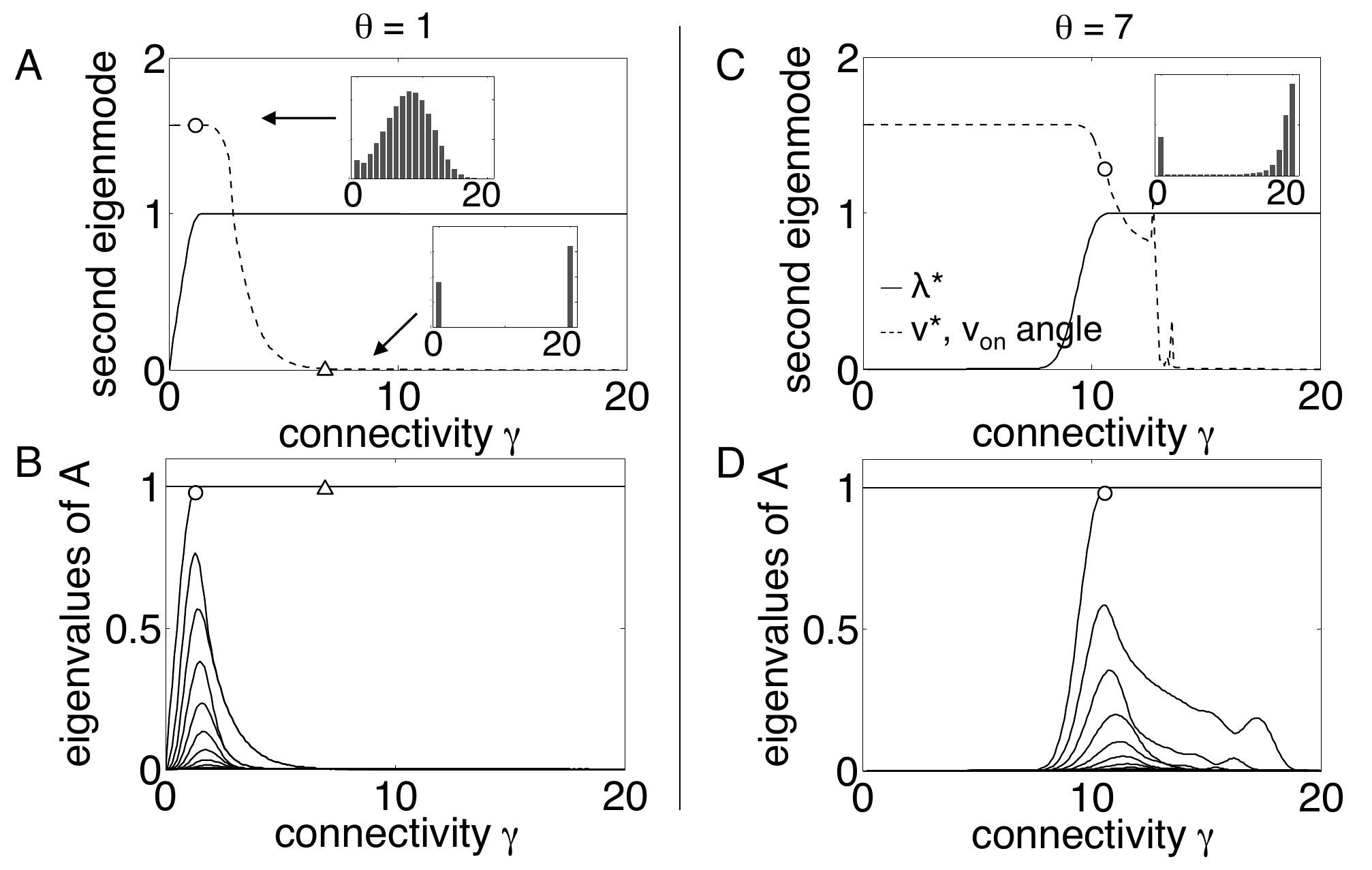}
\end{center}
\caption{Spectral analysis of the mean transition matrix for networks with (A, B) $\theta = 1$ and (C, D) $\theta = 7$. (A, C) The second largest eigenvalue $\lambda^*$ (solid line) effectively converges to $1$ while the angle between $v^*$ and the vector corresponding to full synchrony $v_{\text{on}}$ (dashed line) maintains significant value for a range of $\gamma$, indicating that the second eigenmode is both persistent and far from bimodal. This is also illustrated by the insets, which show typical histograms on the line quasi-attractor either for an intermediate value of $\gamma$ (circle markers on dashed line in A, C and on the second dominant eigenvalue in B, D) or when $\gamma$ is too high to support broadly distributed eigenmodes, resulting in bimodal distributions (triangular marker in A, B). (B, D) Also at the emergence of the line quasi-attractor (circle markers), all eigenvalues are near-maximal compared to their values over the entire connectivity range.}
\label{spectral}
\end{figure}

\begin{figure}[h!]
\hfill
\begin{center}
\includegraphics[width = 5.8 in] {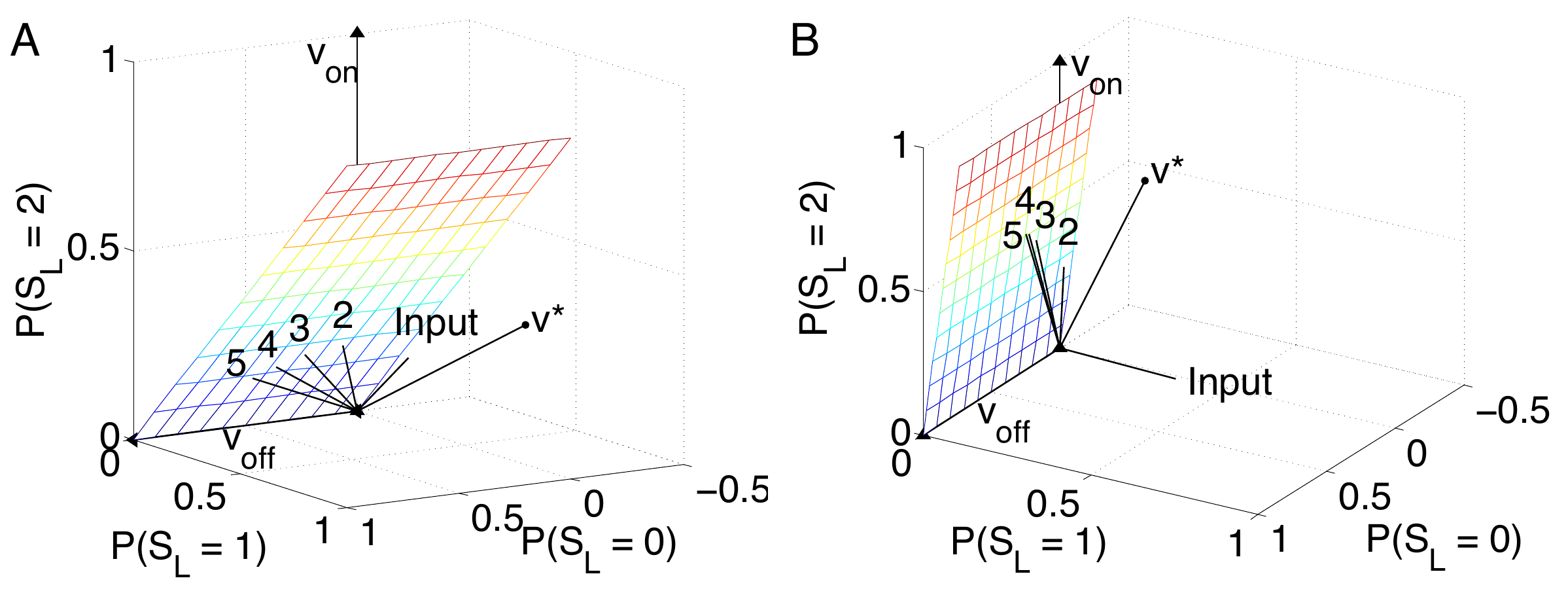}
\end{center}
\caption{Geometrical interpretation of spike count propagation for $N = 2$, $\theta = 1$. The spike count histogram evolves through layers according to the mean-field model via matrix multiplication, affecting rotation in the space of spike count probability distributions. (A) For low connectivity strengths, the input distribution quickly converges to the plane spanned by the first two eigenmodes (the rainbow plane, although histograms will be constrained to the line embedded in the plane satisfying $\sum_i P_i = 1$) as the network encodes the signal.  The distribution then slowly converges to the true stationary state $v_{\text{off}}$, and the signal decays. In this example convergence to quiescence occurs in only a few layers. (B) For high connectivity levels, activity persists through deeper layers, but the line quasi-attractor has rotated closer to the space of bimodal distributions spanned by $v_{\text{on}}$ and $v_{\text{off}}$. The ideal network lies between these two figures. See text for a more detailed discussion.}
\label{geom}
\end{figure}

We analyze the eigenstructure of $A$ through a combination of mathematical analysis and computational argument. First, it can be proven that $A$ has one unique stationary state corresponding to all neurons being quiescent: $v_{\text{off}} = (1,0,\dots,0)$ (Proposition 1 in Appendix). Second, if $A$ is well-behaved in the sense that its eigenvectors have limits as $\gamma \to N$ (an assumption that is supported by numerics, see Figure~\ref{spectral}AC), then the second largest eigenvalue $\lambda^*$ of $A$ converges to 1 as $\gamma \to N$, indicating the emergence of an additional dimension of persistent activity. The catch, however, is that the corresponding eigenvector $v^*$ converges to a vector in the subspace of bimodal or synchronous distributions -- that is, to the span of $v_{\text{off}}$ and $v_{\text{on}} $ where $v_{\text{on}}=(0,...,0,1)$ corresponds to all neurons firing (Proposition 2 in Appendix). All other eigenmodes must converge to 0 as $\gamma \to N$. So despite the emergence of this extra persistent dimension, activity becomes synchronous as connectivity strength increases. Ideally, what we want is for $\lambda^*$ to be practically 1 yet for the span of $v^*$ and $v_\text{off}$ to be far from the plane of bimodal distributions.

Intriguingly, numerical calculations reveal that this does occur for an intermediate level of connectivity $\gamma_{\text{eig}}$ (Figure~\ref{spectral}AC), implying the emergence of a plane spanned by the two principal eigenmodes $v^*$ and $v_{\text{off}}$ that, due to increased persistence, effectively acts as an attractor in sufficiently shallow layers: because of this, we will call this plane \emph{quasi-attracting}. Once the vectors are normalized to represent probability distributions, this means that at $\gamma_{\text{eig}}$, there exists a line quasi-attractor that is far from the space of bimodal distributions, and hence that the network can support broadly distributed, incompletely synchronized firing states. At this same intermediate $\gamma_{\text{eig}}$, we also observe significant values of all eigenmodes (Figure~\ref{spectral}BD), showing further persistent activity contributed by other eigenmodes, at least for the initial network layers.

We pause to give a geometrical view of the mean-field dynamics described above. This is illustrated in Figure~\ref{geom} for $N = 2$, although the following description holds for arbitrary $N$. Consider the $(N+1)$-dimensional space of the spike count probability distribution at a layer. Starting with any input probability vector $P_{\text{input}}$, the layer-to-layer mean-field dynamics of the network can be visualized as iterated rotations of the input vector $P_{\text{input}}$ in the space of spike-count distributions, constrained to the simplex $\sum_i |P_i| =1$. In Figure~\ref{geom}, repeated applications of $A$ are enumerated. In the first few iterations, the spike count distribution converges towards the line quasi-attractor spanned by $v_{\text{off}}$ and $v^*$ as smaller eigenmodes decay.  This represents rapid encoding of the input distribution. Eventually, the system drifts to the stationary state where all neurons are silent, $v_{\text{off}}$.  This of course represents a final state in which the network has ``forgotten" the input. If $\gamma < \gamma_{\text{eig}}$, then the convergence to $v_{\text{off}}$ happens within a few layers (as in Figure~\ref{geom}A). When $\gamma > \gamma_{\text{eig}}$, although activity persists through many layers as expected, the line quasi-attractor has rotated nearer to the span of $v_{\text{on}}$ and $v_{\text{off}}$, so that the persistent activity is nearly synchronous (see Figure~\ref{geom}B). It is only when $\gamma \approx \gamma_\text{eig}$ that activity is persistent while resisting synchrony. In this sense, $\gamma_{\text{eig}}$ represents the existence of a persistent mode of activity characterized by a balance of principal eigenmodes that are broadly distributed, avoiding firing patterns being limited to  synchrony or quiescence.  In fact, as we will explore in the following section, $\gamma \approx \gamma_{\text{eig}}$ also predicts further interesting statistical features of network responses.


\section{Statistical structure of network responses}

Next we will quantify the statistical features of the network responses over the range of connectivity strengths and threshold levels. First, when do responses show that neurons in a given layer fire with {\it higher-order correlations} -- that is, in a way that cannot be predicted from their pairwise spike correlations alone?  Beyond their basic role in characterizing the degree of coordinated spiking in networks~\citep{Shlens, Schneidman, Martignon, Staude}, higher-order correlations have been shown to be necessary to produce broad distributions spiking activity~\citep{Amari} (for recent applications, see~\citet{Macke, Yu}), and to significantly impact coding of stimuli~\citep{Ganmor,Montani}. 

To calculate the extent of higher-order correlations, we utilize maximum entropy models \citep{Shlens, Schneidman, Jaynes}. The pairwise maximum entropy fit of a probability distribution is defined as the distribution that has maximal entropy while being constrained to match the first and second moments of the true distribution.  Thus, this fit makes the fewest additional assumptions on the structure of the probability distribution -- any additional structure is attributed to higher-order correlations. For the permutation-symmetric networks at hand, the pairwise maximum entropy distribution is given by:
\begin{equation*} P^{\text{ME}}(n) = \frac{1}{Z} \exp \{ \lambda_1 n + \lambda_2 n^2  \} \end{equation*} 
where $Z$ is a normalizing factor and $\lambda_1$, $\lambda_2$ are the parameters chosen to match the first two moments. To quantify the effect of higher-order correlations present in the system, we compute the stimulus-averaged Jensen-Shannon (JS) divergence between the true distribution and its maximum entropy fit:  
\begin{align*}D_{JS}(P_L,P_L^{ME}) =& \frac{1}{2} \sum_{m=0}^N P_L(m) \log_2 \left( \frac{P_L(m)}{\frac{1}{2}(P_L(m)+P_L^{ME}(m))} \right) \\ &+\frac{1}{2} \sum_{m=0}^N P_L^{ME}(m) \log_2 \left( \frac{P_L^{ME}(m)}{\frac{1}{2}(P_L(m)+P_L^{ME}(m)} \right).
\end{align*}
This quantity assumes values $0 \leq D_{JS}(P_L,P_L^{ME}) \leq 1$; it can be thought of as a symmetrized version of the Kullback-Leibler divergence. 

\begin{figure}[h!]
\hfill
\begin{center}
\includegraphics[width = 5.9 in] {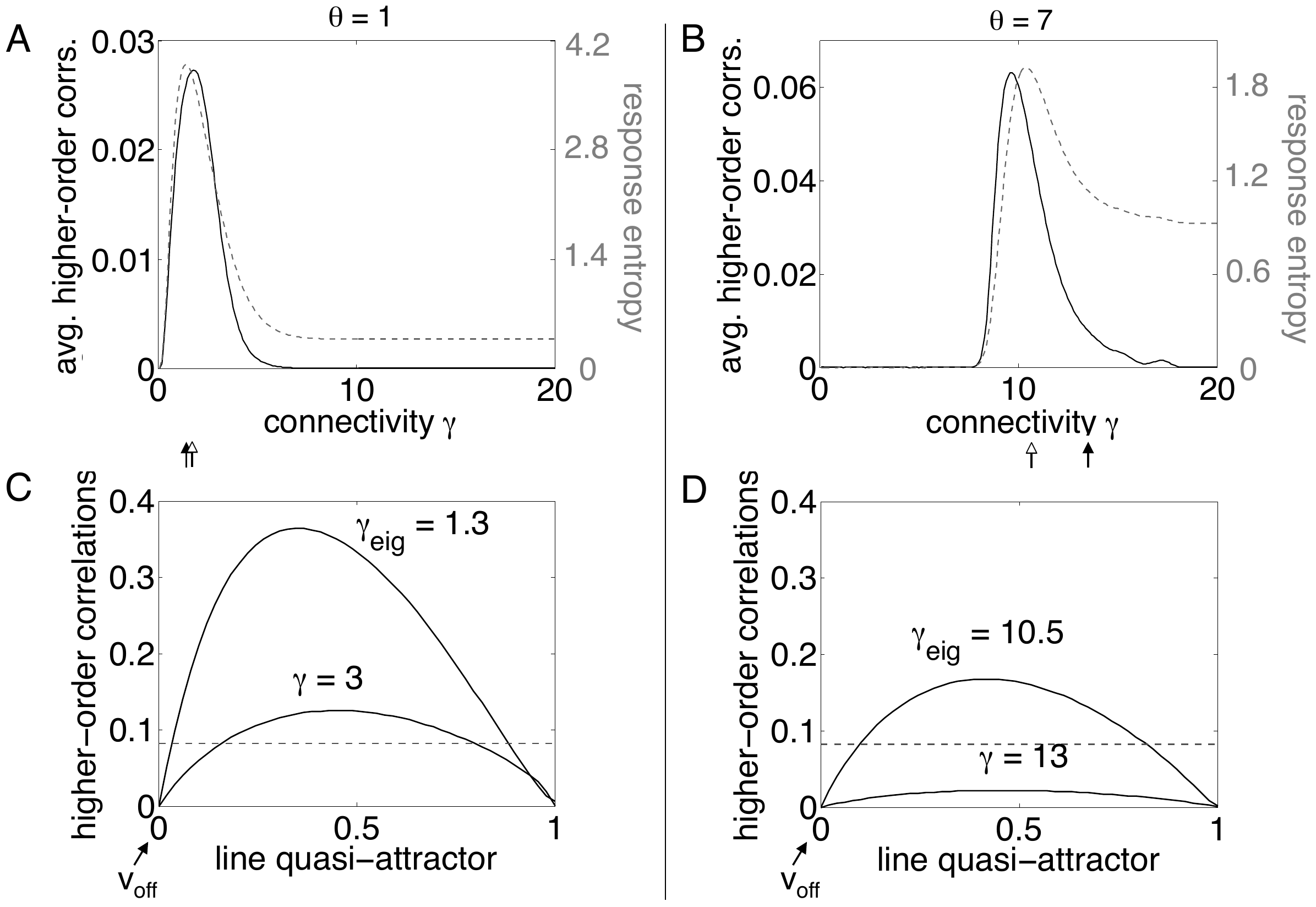}
\end{center}
\caption{Statistical features of network responses for (A, B) $\theta = 1$ and (C, D) $\theta = 7$. (A, C) Response entropy (dashed grey line), and stimulus-averaged higher-order correlations (solid black line) plotted as a function of $\gamma$. Also shown are $\gamma_{\text{eig}}$ (hollow arrow below panel) and $\gamma_{\text{obs}}$ (solid arrow). (A) When $\theta = 1$, peaks in both curves line up with $\gamma_\text{eig}$, as does $\gamma_\text{obs}$  (arrows offset for visibility). (B) For higher threshold networks, $\gamma_{\text{obs}}$ doesn't align with $\gamma_{\text{eig}}$ or other assays (see Section 6). (C, D) Higher-order correlations of spike count histograms along the line quasi-attractor (solid lines; axis parameterizes distributions along the line quasi-attractor starting at $v_\text{off}$). Compare with higher-correlations averaged over the entire space of histograms (dashed lines).}
\label{stat}
\end{figure}

Recall that each neuron in the first layer is independently stimulated so that their firing is a Bernoulli trial, so no correlations are injected into the network. All correlations, pairwise and higher-order alike, emerge solely from the network interactions. We computed the JS divergence between spike count distributions distributions at layer $5$ and their pairwise maximum entropy conditioned on input rate, then average this over all possible stimuli. Through this assessment, we note significant higher-order correlations already by the fifth layer at $\gamma_{\text{eig}}$ (Figure~\ref{stat}AB, solid line). 

How can we understand how these higher-order correlations arise?  We next show that they can be predicted from the spectral analysis of the previous section, without the need for simulation.  Figure~\ref{stat}CD plots the JS divergence between the spike count histograms on the line quasi-attractor and their pairwise maximum entropy fit.  Here, we plot this quantity as a function of their position along the line, parameterized so that $v_\text{off}$ is at position 0. This can be compared to an average JS divergence of approximately 0.08 (dashed lines; calculated by averaging over 10,000 random sample distributions so that the mean had converged) over the entire space of possible response histograms. In particular, the eigenvectors at $\gamma_\text{eig}$ produce significantly larger higher-order correlations than the average. This is because at this level of connectivity, the response distribution is a mixture of two distributions: a large component of quiescent neurons corresponding to $v_{\text{off}}$, and a broader component corresponding to the contribution of $v^*$. As $\theta$ increases, the level of higher-order correlations decreases on the line quasi-attractor, as higher thresholds reduce the breadth of $v^*$.

The second statistical feature of note is the response entropy of the spike count distribution:
\begin{equation*} H(P(S_L))=\sum_{n=0}^N P(S_L=n) \log_2 P(S_L=n). \end{equation*}
Larger response entropies indicate broader response distributions. The response entropy at the 5th layer peaks at $\gamma_{\text{eig}}$, indicating that the emergence of the line quasi-attractor corresponds to the broadest distribution of activity across all values of $\gamma$ (Figure~\ref{stat}, dashed grey line). However, the peak response entropy decreases for higher values of $\theta$; this is the result of the fact that as $\theta$ increases, $v^*$ produces less broad response distributions due to the high threshold and hence the silencing of weak inputs, preventing them from eliciting any firing in the subsequent layer.

In sum, we have shown that at $\gamma_{\text{eig}}$, networks display maximal response entropy and significant contributions from higher-order correlations, directly because of the contribution of other eigenmodes at that level of connectivity.


\section{Combining neutral stability and broad response distributions}

In order to maintain averaged levels of activity without succumbing to synchrony, a network must simultaneously satisfy two criteria.  The first is that it must be able to preserve averaged firing rates from layer to layer without succumbing either to runaway excitation and maximal firing rates in deeper layers, or to decaying network activity.  Second, a network must exhibit a broad spike count distribution at each layer in order to prevent the buildup of correlations and synchrony \citep{Kumar, Reyes, Litvak}.  We refer to these properties, taken together, as  \emph{asynchronous rate coding}.

For which parameter regimes can such {asynchronous} rate coding occur? To quantify this we need an assay that captures how well networks are able to propagate broad response distributions from one layer to the next.  We base this on the propagation of binomial spike count distributions, as these correspond to fully independent activity in each layer. Specifically, we define the spike count JS divergence $D$ to be the JS divergence between the binomial input  distribution $P_1$ and the $L$th layer spike count distribution $P_L$ averaged over all stimuli $S$:
\begin{equation} D(\gamma,\theta)=\frac{1}{N+1}\sum_{n=0}^N D_{JS}(P(S_L|S=n),P(S_1|S=n)). \end{equation}
Networks will exhibit good performance, measured by low values $D$, when they maintain the broad (independent) spike count distribution and the averaged firing rate that occurs in the first layer. 

\begin{figure}[h!]
\hfill
\begin{center}
\includegraphics[width = 4 in] {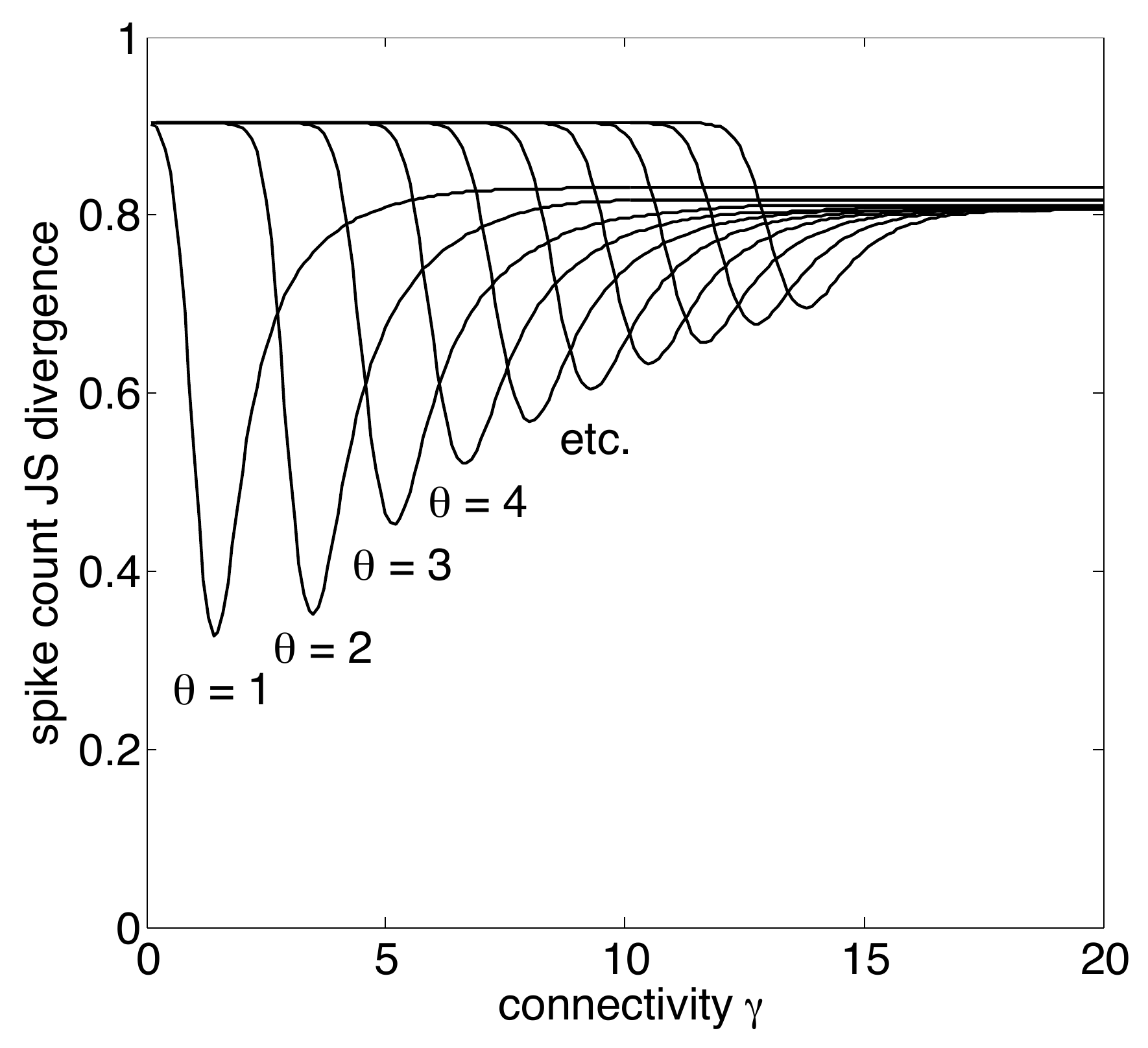}
\end{center}
\caption{Spike count JS divergence plotted as a function of $\gamma$ for increasing $\theta$. Optimal performance for each threshold (the minimum value of the curve) occurs near $\gamma_{\text{eig}}$.}
\label{djs}
\end{figure}

Plots of the spike count JS divergence over $\gamma$ are shown in Figure~\ref{djs} for increasing values of the spike threshold $\theta$. For each fixed $\theta$, there is an optimal, intermediate value of $\gamma$ at which networks are best able to satisfy both of our criteria. However, as threshold level increases, the best value of the spike count JS divergence also increases, showing that high-threshold networks fail to produce asynchronous rate coding.

This failure follows from the requirements of neutral dynamics and broad response distributions described in previous sections. First, from Section 3, $\gamma_{\text{obs}}$ captures the first criterion of complex signal coding outlined above: that is, networks demonstrate neutral stability and average one-to-one rate transmission when they average a branching ratio of $\sigma \approx 1$. On the other hand, Section 4 shows that $\gamma_{\text{eig}}$ reflects when the network supports persistent, broad response distributions, providing an assay of the second criterion. Complex signal propagation can therefore only occur in these systems when $\gamma_{\text{obs}} \approx \gamma_{\text{eig}}$. Comparing Figure~\ref{spectral} with the previous Monte Carlo simulations in Figure~\ref{schematic}BC reveals both criteria can indeed be simultaneously satisfied when few inputs are required to cause a spike, but a gap between these required values of connectivity $\gamma$ appears with increasing $\theta$. To be precise, for $N = 20$, $\theta = 1$, Monte Carlo simulations and spectral analysis both yield $\gamma_{\text{obs}} \approx \gamma_{\text{eig}} \approx 1.3$. When $\theta = 7$, however, simulations show $\gamma_{\text{obs}} \approx 13.75$ yet $\gamma_{\text{eig}} \approx 10.5$. As shown through the eigenstructure, at $\gamma_{\text{obs}}$, only bimodal activity is supported after a few layers. In fact, because of the inevitable synchrony in deep layers, optimal performance under the JS divergence tends to fall nearer to $\gamma_{\text{eig}}$ than to $\gamma_{\text{obs}}$. Networks of high-threshold neurons are therefore unable to simultaneously satisfy both requirements of complex signal propagation outlined at the beginning of this section.

Intuitively, the reason that $\gamma_{\text{obs}} > \gamma_{\text{eig}}$ is that networks with high-threshold  neurons reject inputs of low firing rate, so that when $\theta$ is large there is an increased likelihood that connectivity structure and stochasticity will conspire to silence all activity in the next layer. Geometrically speaking, as $\theta$ increases so does the nullity of $A$, resulting in a larger and larger subspace that trajectories must avoid lest they risk susceptibility to network quiescence; in order to reach an average of one-to-one rate transmission, it is necessary to provide a buffer for the coding subspace from the nullspace by inflating the connectivity into the regime of bimodality. 

Also of practical importance is the question of robustness to parameters. The delicate nature of $\gamma_{\text{eig}}$ and $\gamma_{\text{obs}}$ constrains networks that produce asynchronous rate coding to finely-tuned connectivity strengths; one requires that the branching ratio lie at a critical value $\sigma \approx 1$, while the other relies on a precise balance between persistent yet broadly supported eigenmodes. This sensitivity is reflected in the sharp troughs in the JS divergence (Figure~\ref{djs}); for even higher $N$, these troughs become even sharper and robustness is a more important goal to obtain. As we will see in the next section, however, this sensitivity can be mitigated by adding an inhibitory population to each layer.

To summarize results thus far, we evaluate networks on two criteria: $\sigma \approx 1$ and broad response distributions. Low-threshold networks can always satisfy broad response distributions and maintained average rate transmission at the same $\gamma$. High-threshold networks are able to somewhat support broad distributions, although the preserved aspects of network responses and their lower values of response entropy indicate less broad distributions as compared to their low-threshold counterparts. They also can satisfy $\sigma \approx 1$, however this is due to averaging: because of the increasing nullity of the mean transition matrix, these networks cannot propagate weak input stimuli, so they must overcompensate by inflating $\gamma$. Because of this, no high-threshold network of a fixed $\gamma$ can simultaneously both criteria, and hence they cannot propagate rates asynchronously through layers. This appears to be a significant limitation for high-threshold networks -- and, importantly, for many biological neural networks, in which many inputs are required to elicit a spike. In the following sections we will incorporate additional biophysical features -- inhibition and noise -- and study whether this provides a resolution so that high-threshold networks can support persistent, broadly distributed activity.


\begin{figure}[h!]
\hfill
\begin{center}
\includegraphics[width = 5.8 in] {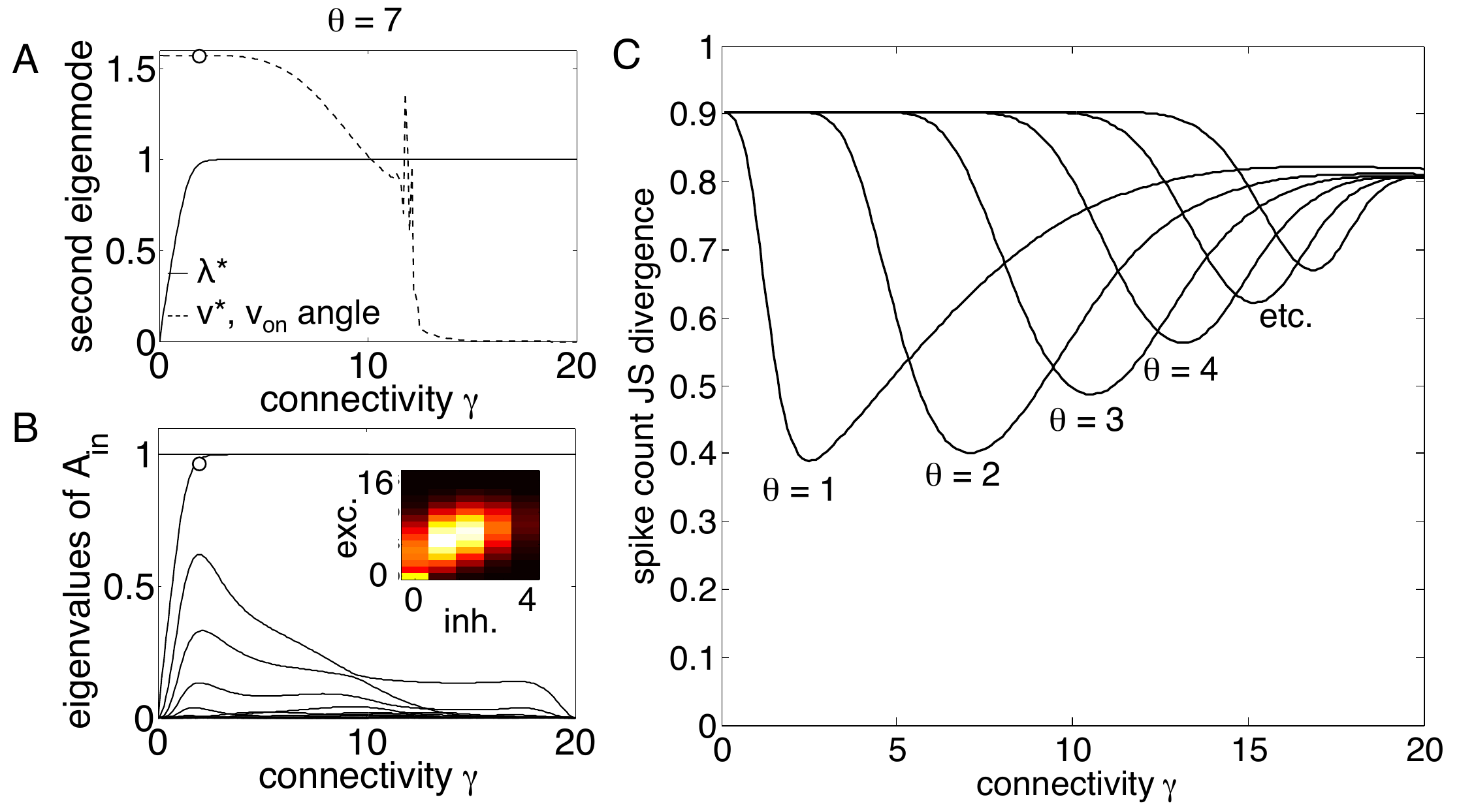}
\end{center}
\caption{Excitatory-inhibitory networks display increased robustness, $N_E = 16$ and $N_I = 4$. (A) The second largest eigenvalue $\lambda^*$ and the angle between $v^*$ and $v_{\text{on}}$ (dashed line) overlap over a larger parameter space, indicating robustness at $\gamma_{\text{eig}}$; (B) similarly, all eigenvalues $A$ of have broader peaks. Inset shows a typical broadly distributed histogram at $\gamma_{\text{eig}}$ (indicated by the marker in A, B). (C) The spike count JS divergence has a wider minimum  for all values of $\theta$, showing that inhbition also allows for more robust asynchronous rate propagation.}
\label{inhib}
\end{figure}

\section{Excitatory-inhibitory networks display increased robustness}

How can asynchronous rate propagation emerge in high-threshold networks? Intuitively, we might expect an added inhibitory population to prevent runaway excitation and saturation of firing rates to high values, thus preventing synchrony. To test this, we added an inhibitory population of $N_I$ neurons to each layer of $N_E = N - N_I$ excitatory neurons, and further impose $N_E-N_I>\theta$ (otherwise no activity could be transmitted due to the homogeneity in network connectivity -- even if only the excitatory population is active in layer $L$, the random connectivity imposed will cause the same proportion of the excitatory and inhibitory populations in layer $L+1$ to fire). Network parameters are assumed to be homogenous among the inhibitory and excitatory populations. Because of this assumption, it is straightforward to calculate the new, four-dimensional mean-field transition matrix $A_{\text{in}}$: 
\begin{align*}
P(S_L^E=m_E, S_L^I=&m_I | S_{l-1}^E=n_E, S_{l-1}^I=n_I) = \\
 & {N_E \choose m_E} q_{n_E,n_I}^{m_E} (1-q_{n_E,n_I})^{N_E-m_E} \times {N_I \choose m_I} q_{n_E,n_I}^{m_I} (1-q_{n_E,n_I})^{N_I-m_I},
\end{align*}
where $S^i_L$ is the number of cells spiking in the excitatory ($i=E$) or inhibitory ($i=I$) population at the $L$th layer, and $q_{n_E,n_I}$ is the probability that a downstream neuron spikes given $n_E$ spiking excitatory neurons and $n_I$ spiking inhibitory neurons in the upstream layer:
\begin{equation*} q_{n_E,n_I}=\sum_{k_E=\theta}^{n_E} \sum_{k_I=0}^{\min(n_I,k_E-\theta)} {n_E \choose k_E}\left(\frac{\gamma}{N}\right)^{k_E} \left( 1-\frac{\gamma}{N} \right)^{n_E-k_E} \times  {n_I \choose k_I}\left(\frac{\gamma}{N}\right)^{k_I} \left( 1-\frac{\gamma}{N} \right)^{n_I-k_I}. \end{equation*}
The binomial input distributions now take the following form:
\begin{align*} P(S_1^E&=m_E, S_1^I=m_I | S=n) \\
&= {N_E \choose m_E} \left(\frac{n}{N}\right)^{m_E}\left(1-\frac{n}{N}\right)^{N_E-m_E} \times {N_I \choose m_I} \left(\frac{n}{N}\right)^{m_I}\left(1-\frac{n}{N}\right)^{N_I-m_I}. \end{align*}
 
The expression for the transition matrix for the excitatory-inhibitory networks has a similar form to that of the purely excitatory networks, so the eigenstructure of $A_{\text{in}}$ is similar to that of $A$: it has a unique stationary state corresponding to all neurons being quiescent, and as $\gamma \to N$, the second largest eigenvalue converges to 1 and its eigenvector corresponds to bimodality (Proposition 3 in Appendix). There also is an intermediate state of connectivity at which $\lambda^* \approx 1$ and $v^*$ is far from bimodal (Figure~\ref{inhib}AB). Here we consider $\theta = 7$, as well as $N_E = 16$, $N_I = 4$ to simulate $\sim$20\% inhibition, as typically used in, for example, cortical modeling (cf.~\cite{Braitenberg}).\footnote{We emphasize that results in this section are not particular to these specific choices of $N_I$ and $N_E$. As long as $N_E-N_I>\theta$, the intermediate $\gamma_{\text{eig}}$ producing broad, persistent distributions will continue to exist. Other results regarding robustness, and limitations on asynchronous rate propagation high $\theta$, also continue to hold.} This yields $\gamma_{\text{eig}} \approx 16.9$. However, according to Monte Carlo simulations, the branching ratio is always less than 1 for all $\gamma < N$. Firing rates thus fail to be maintained in this network, as reflected in the spike count JS divergence in Figure~\ref{inhib}C. The reason is that $A_{\text{in}}$ is so structurally similar to $A$: as in the purely excitatory networks, the high threshold still rejects weak inputs and sends them to the stationary state of quiescence, $v_{\text{off}}$. This is in agreement with  \citet{Reyes}, who found that adding a homogenous inhibitory population to each layer does not help networks avoid synchrony.

Inhibition does, however, increase the robustness of JS divergence to perturbations in connectivity strength $\gamma$.  Specifically, the troughs of minimal JS divergence widen compared to those of purely excitatory networks (Figure~\ref{inhib}C). This is reflected as well in the eigenstructure: the intermediate state of persistent, broadly distributed distributions is now stretched to cover a wider range of $\gamma$ (Figure~\ref{inhib}AB). This robustness further grows as the size of the inhibitory population is increased, so long as $N_E - N_I >\theta$ (results not shown). Intuition for this effect can be obtained by comparing to the purely excitatory case. Suppose a typical neuron in this case has $n$ inputs. To produce a broad range of responses and avoid either too many inputs (resulting in maximal firing rates) or too few (resulting in quiescence), $n$ must hover near some critical value that depends on particular choice of parameters. Now consider an excitatory-inhibitory network: then the typical neuron has $(1-N_I/N_E)n$ net inputs when taking into account inhibition. Since $1-N_I/N_E<1$, this slope is shallower than that for purely excitatory networks, so the networks are more robust to chance perturbations around the critical value of inputs, and hence to connectivity strength.

Increased robustness to connectivity parameters in the presence of inhibition is interesting as it addresses a major concern regarding the plausibility of dynamics at critical transition values of connectivity (as discussed in the previous section). In sum, inhibition may help resolve the need for fine-tuning by enhancing robustness to fluctuations in network connectivity.


\section{Impact of background noise}

The next attempt to recover asynchronous rate propagation follows from~\citet{van Rossum}, in which a noisy background current was shown to enhance the preservation of firing rates in feedforward networks of integrate-and-fire neurons; see also ~\citet{Nowotny}, \citet{Reyes}, and \citet{Litvak}. We inject background noise in the form of independent, zero-mean Gaussian independent noise current $\chi$ to each neuron, $\chi \sim \mathcal{N}(0,\sigma_{\chi}^2)$. This transforms the heaviside-like thresholding into a smoother, sigmoidal operation. The probability that a neuron will spike given $n$ cells firing in the upstream layer is now
\begin{equation*} q_n = \int_{-\infty}^{\infty} Pr(I>\theta-x)Pr(\chi=x)dx,\end{equation*}
where $I$ is the synaptic input from the upstream layer without the additional noise component. If $x>\theta$ then the neuron fires with probability $1$ because the noise alone is enough to elicit a spike. If $x<\theta-n$, then the neuron can never fire as even the addition of all upstream neurons delivering input would be insufficient to cross threshold. We can then rewrite $q_n$ as:
\begin{align*}
q_n =& \int_{\theta-n}^{\theta} \frac{1}{\sqrt{2 \pi}\sigma_{\chi}} \exp\left(-\frac{x^2}{2\sigma_{\chi}^2}\right) \left[ \sum_{k=\lceil \theta-x \rceil}^n {n \choose k} \left( \frac{\gamma}{N} \right)^k \left(1-\frac{\gamma}{N} \right)^{n-k}\right] dx\\
&+\int_{\theta}^{\infty} \frac{1}{\sqrt{2 \pi}\sigma_{\chi}} \exp\left(-\frac{x^2}{2\sigma_{\chi}^2}\right) dx.
\end{align*}
\citet{Nowotny} consider a similar expression (Equation 7 in their paper), although both the exact form of their expression, and their conclusion that it has little effect on the transition matrix, differ from ours. We denote by $A_\text{noisy}$ the transition matrix describing these networks with added noise, generated by the new $q_n$.

\begin{figure}[h!]
\hfill
\begin{center}
\includegraphics[width = 4.2in] {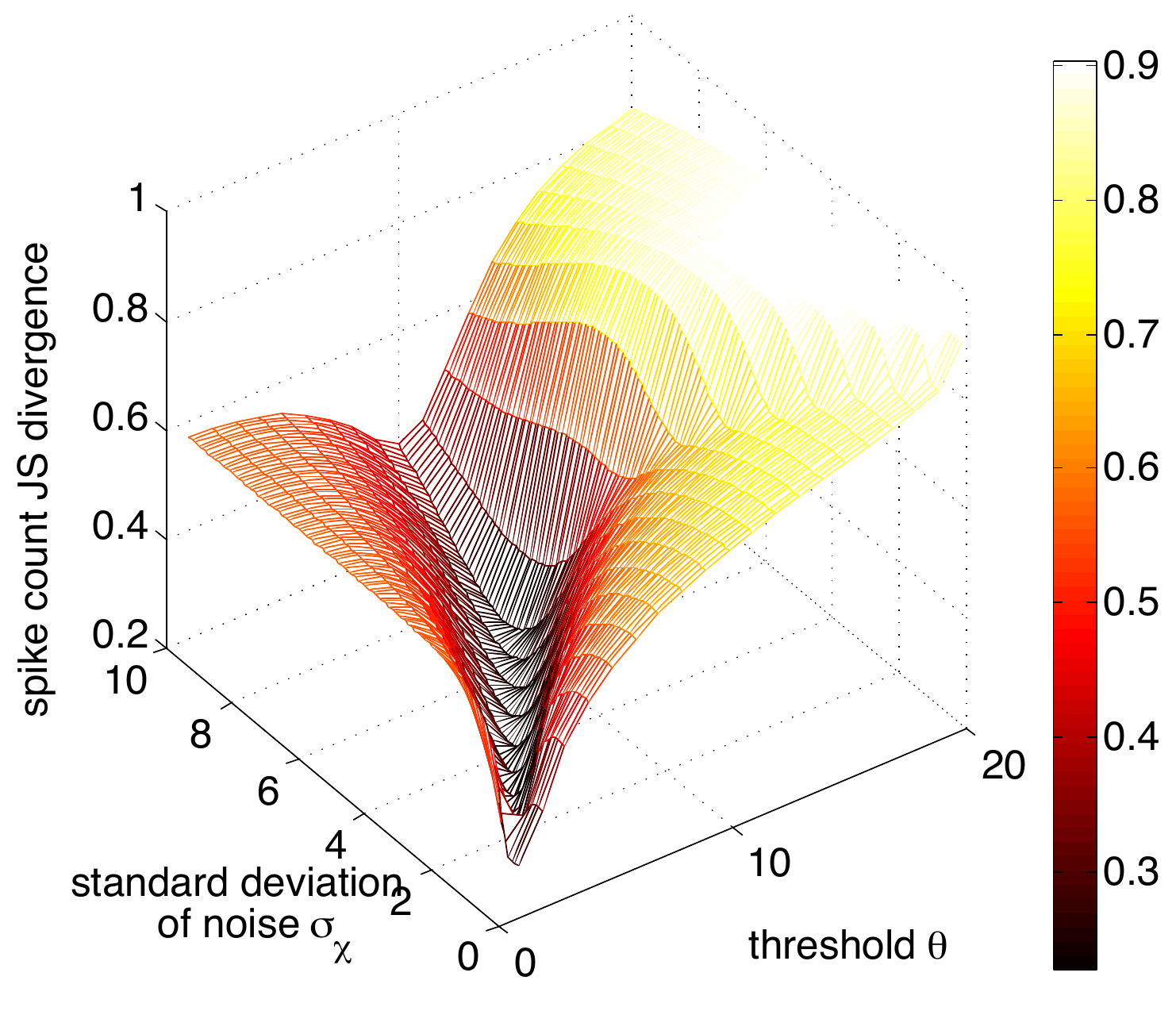}
\end{center}
\caption{Impact of noise on input propagation.  Surface shows spike count JS divergence as a function of $\theta$ and the standard deviation of noise added to each neuron, $\sigma_{\chi}$. For each $\theta$, there is a $\sigma_{\chi}$ that optimizes asynchronous rate propagation.  For $\theta < 10$ the relationship between $\theta$ and optimal $\sigma_{\chi}$ is linear.}
\label{noise3d}
\end{figure}

Figure~\ref{noise3d} plots the spike count JS divergence (Equation 1) as a function of $\theta$ and $\sigma_{\chi}$.  The main result is that adding noise produces lower values of JS divergence -- and thus more consistent propagation of asynchronous inputs -- at larger values of $\theta$.  This result agrees with the findings of  \citet{van Rossum} (cf. their Figure 2B and see Appendix for further comparisons with this study).  Our result is also in agreement with \cite{Reyes}, who finds that adding white noise as a background current reduces the amount of synchrony present in networks. 

\begin{figure}[h!]
\hfill
\begin{center}
\includegraphics[width = 5.8 in] {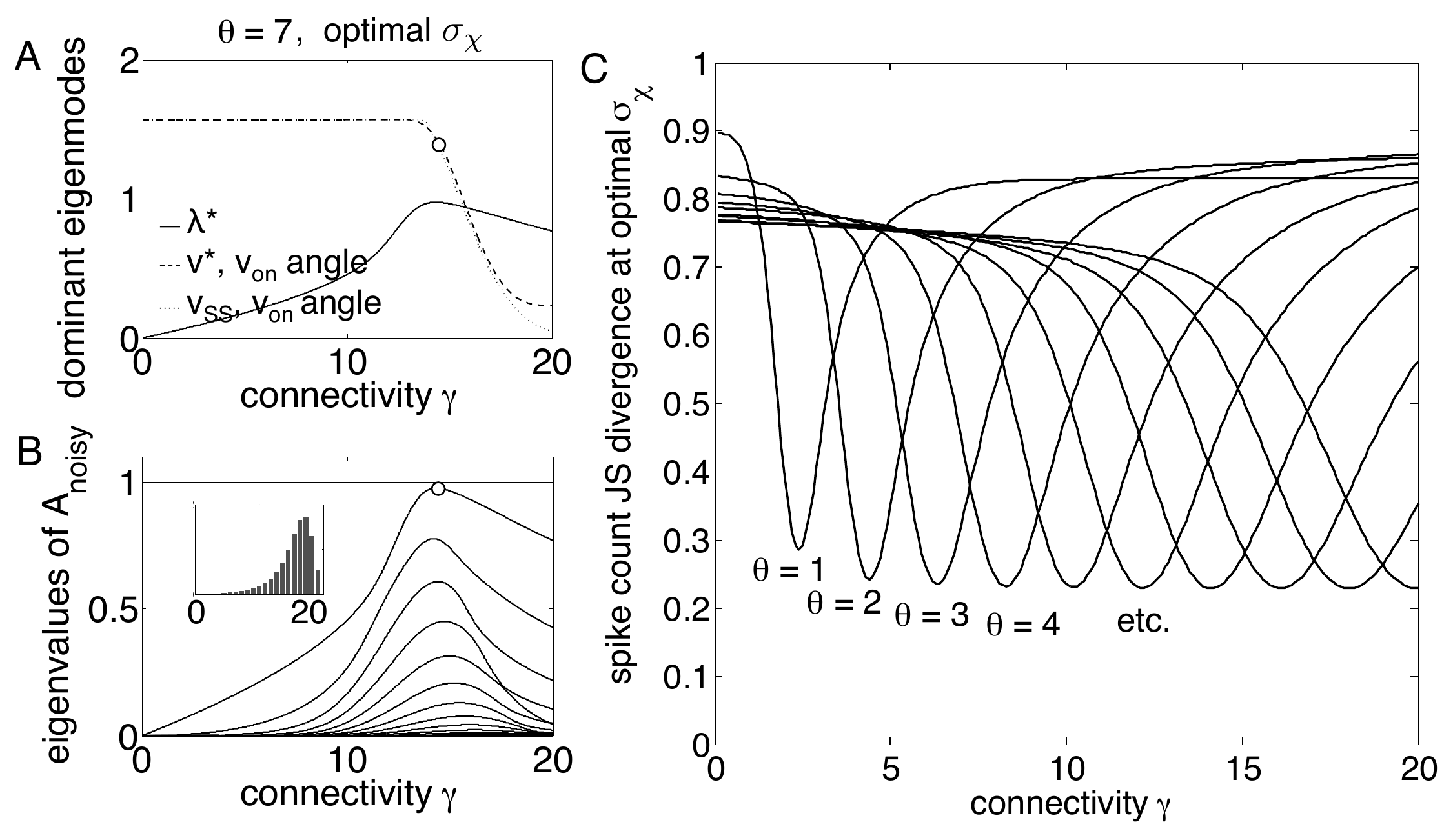}
\end{center}
\caption{Properties of noisy networks at optimal background noise levels.  (A) The second largest eigenvalue $\lambda^*$ peaks very close to 1 at an intermediate $\gamma_{\text{eig}}$. The angle between $v^*$ and $v_{\text{on}}$ (dashed line) and between $v_\text{SS}$ and $v_{\text{on}}$ (dotted line) have large values at $\gamma_{\text{eig}}$. (B) At this same value $\gamma_{\text{eig}}$, all eigenmodes have significant contribution. Inset shows a typical broadly distributed histogram at $\gamma_{\text{eig}}$ (indicated by the marker in A, B). (C) The spike count JS divergence, taking the optimal value of $\sigma_{\chi}$ for each $\theta$. With optimal noise values added, asynchronous rate propagation is dramatically improved for high-threshold networks.}
\label{noise}
\end{figure}

For each of the threshold values $\theta$ shown, there is an optimal $\sigma_{\chi}$ for asynchronous rate propagation (i.e., that minimizes the JS divergence).  This amount of noise gives spontaneous firing rates of less than 12\%, as measured by the probability $Pr(\chi>\theta)$. For the remainder of this section, we will take the optimal value of noise for each value of threshold. Figure~\ref{noise}C uses these values to provide another view of optimized JS divergence, which reveals the improvement in comparison with noise-free networks (Figure~\ref{djs}).  Moreover, $\gamma_{\text{obs}}$ and $\gamma_{\text{eig}}$ do coincide in the noisy case, even for high values of $\theta$ (at about 14.25 in for $\theta=7$, branching ratio figure not shown).  

In contrast to the effects of inhibition, the addition of background noise does produce substantial changes in the structure of the transition matrix.  For example, comparing $A_{\text{noisy}}$ with $A$, spontaneous activity is now possible, as $v_{\text{off}}$ is no longer the stationary state. Instead, the stationary state $v_\text{SS}$ is now a function of $\gamma$. In particular, the noise contributes a nonzero probability from transitioning from any state to any other state, so the components of $A_{\text{noisy}}$ are strictly positive. By the Perron-Frobenius theorem, this means the system has a unique stationary state $v_\text{SS}$ whose components are all strictly positive (so it can never be $v_{\text{on}}$ or $v_{\text{off}}$). Computationally we find that the second largest eigenvalue now does not converge to 1 as $\gamma \to N$; it does, however, attain a peak value near $1$ at an intermediate $\gamma_{\text{eig}}$, and at this point $v_\text{SS}$ and $v_{\text{off}}$ are also far from bimodal (Figure~\ref{noise}AB). Thus despite the differences in eigenstructure between $A$ and $A_{\text{noisy}}$, the predominant features that define the existence of a persistent set of broad firing distributions are still apparent: there is an intermediate connectivity level $\gamma_{\text{eig}}$ at which all eigenvalues are significant, the second largest eigenvalue in particular is close to 1, and both the stationary distribution and the second eigenmode are far from bimodal.

\begin{figure}[h!]
\hfill
\begin{center}
\includegraphics[width = 5.8 in] {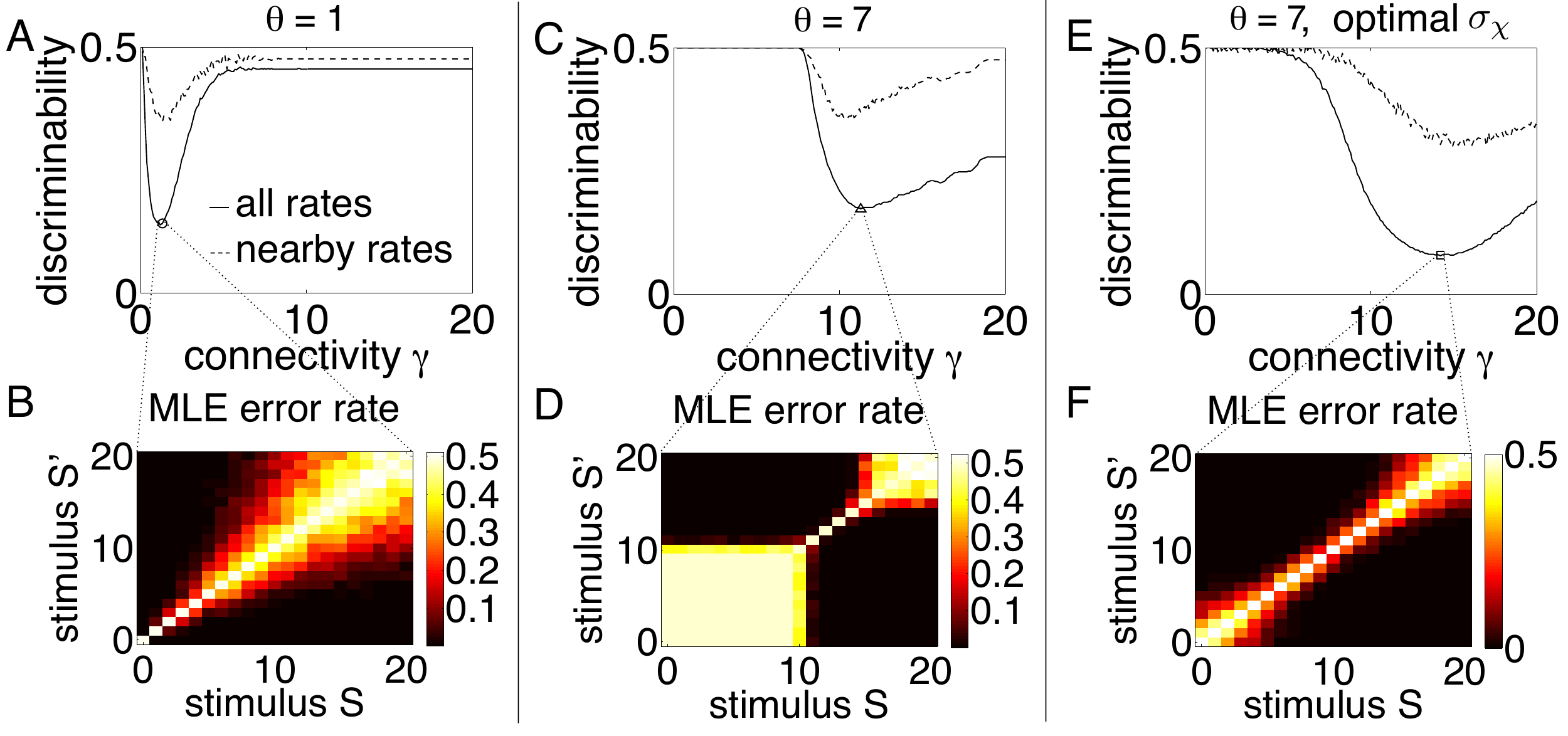} 
\end{center}
\caption{Rate discriminability for noise-free networks with (A, B) $\theta = 1$, and (C, D) $\theta = 7$, as well as (E, F) a network with $\theta = 7$ and optimal noise, $\sigma_{\chi} = 4.2$. (A, C, E) Nearby discriminability (dashed line) and average discriminability (solid line) for $T = 25$ trials plotted as a function of connectivity level $\gamma$. (A) For low-threshold networks, rate discriminability is optimal at $\gamma_{\text{eig}}$. (C) For high-threshold networks, nearby discriminability is best near $\gamma_{\text{eig}}$, but this minimum is shifted for average discriminability. (E) Adding noise improves rate discrimination in high-threshold networks. (B, D, F) Maximum likelihood error rate plotted for every possible pair of input stimuli before averaging. The chosen networks are those that minimize average discriminability, as indicated by the markers in (A, C, E).}
\label{discrim}
\end{figure}

Finally, to put the role of noise to a more demanding test, we test its impact on the capacity of networks to discriminate between different input stimuli. For this, we calculate the \emph{rate discriminability} by measuring the error rate given by the maximum likelihood estimator on $T$ trials.  Specifically, suppose the network produces output spike counts $S_L^1,\dots,S_L^T$ under some fixed input stimulus level $S$. Since the trials are independent, the maximum likelihood estimator (MLE) chooses between two stimuli $S$ and $S'$ by selecting the one that is likelier to result in the given data, via the likelihood ratio:
\begin{equation*} \prod_{j=1}^T \frac{P(S_L^j|S)}{P(S_L^j|S')}. \end{equation*}
If this product is greater than 1 the MLE chooses stimulus $S$; less than 1 and the MLE chooses stimulus $S'$. Assuming $S$ and $S'$ are equally likely {\it a priori}, the error rate is given by 
\begin{equation*} \text{ER}(S,S') = \frac{1}{2}\mathbb{E}\left[ \mathbb{I}\left(\prod_{j=1}^T \frac{P(S_L^j|S)}{P(S_L^j|S')} > 1\right) \middle| S' \right]+\frac{1}{2}\mathbb{E}\left[ \mathbb{I}\left(\prod_{j=1}^T \frac{P(S_L^j|S)}{P(S_L^j|S')} > 1\right) \middle| S \right],\end{equation*}
where the first expectation is taken over the distribution $P(\cdot|S')$ and the second over $P(\cdot|S)$.  This produces an $(N+1)\times(N+1)$ matrix describing the MLE error rate for distinguishing $S$ from $S'$. We then average over either the entries in the entire matrix to give either the average discriminability, or we average the entries in the superdiagonal to give the \emph{nearby} discriminability, i.e, the discriminability between nearby rates.

Figure~\ref{discrim}AC first summarizes discriminability in the absence of noise.  Rate discriminability reaches its minimal value at $\gamma_{\text{eig}} \approx \gamma_{\text{obs}}$ when $\theta = 1$; when $\theta = 7$, the minimal discriminability does not exactly coincide with either $\gamma_{\text{eig}}$ or $\gamma_{\text{obs}}$. A glance at the MLE error rates sans averaging reveals the particular type of computation performed in each case: Figure~\ref{discrim}BD shows the error rates at the values of $\gamma$ that yield the lowest average discriminability, as indicated by the markers in Figure~\ref{discrim}AC. Low-threshold networks are able to accurately discriminate between rates over the entire stimulus space, including nearby rates. High-threshold networks, on the other hand, although able to perfectly distinguish a few rates in a limited intermediate range, cannot at all distinguish between nearby high rates or low rates. Rather, these networks are better suited to classifying input rates into two bins: low and high.

Interestingly, the added background noise promotes better discriminability between rates in high-threshold networks, dropping the minimal level to values even below that of noise-free, low-threshold networks (Figure~\ref{discrim}E). Moreover, the MLE error rates (Figure ~\ref{discrim}F) show a marked improvement in the ability to distinguish between nearby rates at $\gamma_\text{eig}$, as revealed by the tightly banded matrix structure. Not only, then, does noise improve rate propagation in neurons -- it also changes the computation from a coarse-grained classifier to one with more resolution.  This is a specific example of the more general phenomenon of stochastic resonance (see, e.g. \citet{McDonnell,Longtin}).


\section{Discussion}

\subsection*{Summary}

In this paper, we study the transitions in feedforward network dynamics that occur as connectivity strength and firing threshold are varied.  We characterize these transitions via critical branching, neutral stability, higher-order correlations, and broad firing distributions.  After quantifying critical branching by computing the branching ratio, we show that neutral stability (persistence of firing patterns from one network layer to the next), together with statistical properties of the persistent patterns, can be predicted via a spectral analysis of the underlying mean-field transition matrix. Throughout most of the parameter space, persistent activity is restricted to highly bimodal, synchronous responses, as found by \citet{Reyes}, \citet{Nowotny}, and \citet{Litvak}. However, there are ``transition" connectivity levels that yield persistent, broadly-distributed spike count histograms with higher-order correlations and large response entropy.  For low threshold networks this occurs simultaneously with (approximately) critical branching, revealing that such networks are well-suited to transmitting rates without synchronization. On the other hand, high-threshold networks do not produce both critical branching and broad response distributions at the same connectivity strength; when the former is satisfied, these networks tend to produce synchronous responses.

Interestingly, adding further biologically-motivated features increased the robustness of transitions in high-threshold networks.  In particular, simulations and spectral analysis show that including an inhibitory cell population extended the connectivity range that yields asynchronous propagation of inputs.  Adding zero-mean noise to each neuron had a similar effect and also improved the discriminability of inputs, echoing the findings of~\cite{van Rossum} in integrate-and-fire networks.

We conclude that networks with low firing thresholds, or those in which intrinsic noise elevates firing probabilities, exhibit a set of dynamical and statistical signatures associated with ``critical" transitions in network activity.

\subsection*{Connections with the criticality literature}

We now discuss links with the broader literature on criticality, which suggests that the brain may operate at a state characterized by complex dynamics, significant higher-order correlations, and enhanced computational properties.  This is often described as operating on the boundary between ordered and irregular (or chaotic) activity. In particular, such systems can flexibly perform a wide range of operations on time-dependent inputs when their recurrent networks lie near the ``critical" state, which is defined by calculating the expected neutral separation of trajectories using a mean-field model \citep{Bertschinger,Legenstein}.

Along these lines \citet{Beggs} motivates a feedforward model based on array recordings.\footnote{The authors argue that a feedforward model is appropriate in this context as  electrode sites are rarely active more than once during the cascades of neural activity that they study.}  Here, the authors compute the mutual information between the $2^N$ possible binary ``words" at the first and last layers.  Intriguingly, they numerically show  -- for the low threshold case $\theta = 1$ --   that the mutual information is maximized for the same parameters at which critical branching occurs.  Our finding in the averaged, mean-field setting echoes this result.  An interesting extension of our work would be to explain the findings of  \citet{Beggs}  via the spectral properties of the allied layer-to-layer transition matrix between binary words. 

A number of experimental and theoretical studies focus on  \emph{neuronal avalanches} as a signature of critical neural connectivity.  These are cascades of neural activity whose sizes obey a power law distribution \citep{Beggs, Kitzbichler, Hahn, Petermann, Hennig, Mora}. Avalanches have been shown to arise in some neutrally stable models of neural networks \citep{Haldeman}, and thus have been described as a signature of optimal computation. An interesting target for future work would be to extend our mean-field analysis to predict the occurrence of avalanches over multiple network layers, and to  and study their role in encoding stimuli.

\subsection*{Verifying and extending the model}

We imposed a number of simplifications in this paper to achieve analytical tractability.  The most prominent of these is that our neurons are modeled as simple thresholding units with no intrinsic properties or time dependence~\citep{Nowotny}. However, our results  agree those in networks of more realistic neurons \citep{van Rossum, Reyes, Rosenbaum}.  We therefore believe that our findings will prove to be quite general.

Another possible limitation is that the numerical studies presented above utilize a fixed value of $N=20$ neurons.  However, our analytical results on spectral properties of the transition operator are independent of this choice.  Moreover, we verified that our main qualitative results are preserved, e.g., for the larger value $N = 100$ (taking $\theta = 1, 5, 10, 20, 35$); data not shown. In more detail, as with the smaller network, the system at $N=100$ remains well-described by a mean-field transition matrix (in fact, due to the larger population size, it is even better fit). The eigenstructure of these matrices reveals an intermediate $\gamma_\text{eig}$ at which the second dominant eigenmode is both persistent and broadly-distributed, and there is significant contribution from all eigenvalues as well as maximal response entropy. For $\theta = 1$, this value overlaps with $\gamma_\text{obs}$, but as threshold increases, the gap between the two widens; accordingly, the spike count JS divergence increases.  As for the $N=20$ case, while inhibition does continue to increase this range, the optimal performance is not improved. The addition of noise in large networks, however, has similar beneficial effects:  an optimal amount of noise lowers the minimum JS divergence to around 0.32 for high values of $\theta$. This amount of background noise required generates less than 10\% probability of spontaneous firing, similar to that obtained at $N=20$.  However, one difference at $N=100$ is that the optimal performance under the JS divergence metric $D$ is lower:  when $\theta = 1$, the optimal network attains at best a score of 0.58, compared to 0.33 for $N=20$.  Moreover, in the larger network the ``well" in $D$ values near the optimal $\gamma$ value is even narrower, requiring a finer tuning of $\gamma$.  
These findings suggest that, while our findings remain qualitatively similar for larger networks, there may be interesting new phenomena in the continuum limit of large $N$ --  an interesting subject of future study. 

On another note, we focused on only a few of the many metrics of signal propagation and  coding that could be applied to the networks at hand.  We note further results on one of these in the appendix, that used by \citet{van Rossum} to measure propagation of firing rates via trial-to-trial variance of responses in deep layers.  This showed similar results to our measure $D$ of JS divergence between input and output distributions over intermediate firing rates; the two measures showed distinctions at extreme firing rates,  assessing the quiescent or saturating patterns that occur there differently (see appendix).

Finally, we have concentrated solely on networks with a feedforward connectivity structure. However, these networks are equivalent to a synchronously-updated discrete-time network with random recurrent connections (including connections to themselves) under the annealing approximation \citep{Bertschinger}.  Thus, to the extent that these assumptions hold, the results of this paper may also be applied to the evaluation of persistent activity in recurrent networks. 

We close by noting experimental predictions of our work, as could be tested directly in {\it in-vitro} feedforward networks (using the techniques of \citet{Reyes}), or, with the considerations above, could predict dynamics in recurrent systems as well. First, asynchronous rate propagation should become possible when the membrane potentials of neurons are biased upwards (equivalent to decreasing the spike-generation threshold).  Second, this should also occur when sufficient noise is added local to each cell (some white noise has already been shown to reduce synchrony in \citet{Reyes}).   Finally, adding an inhibitory population at each layer should increase the robustness of asynchronous propagation to network connectivity and synaptic strength.

\section*{Acknowledgements}  
We thank John Beggs for valuable insights and discussions.  This research was supported by an NSF Graduate Research Fellowship and an ACRS Fellowship to A.C.G. and by NSF grants DMS-0817649 and DMS-1056125 (CAREER), and by a Career Award at the Scientific Interface from the Burroughs-Wellcome Fund (to E.S.-B.).

\section*{Appendix}

\subsection*{Derivation of mean-field Markov chain}
We outline how to obtain a formula for the mean-field Markov chain given the transition matrix for the original Markov chain in the space of $2^N$ firing patterns. The first step in this derivation is to determine the probability that, given true connectivity matrix $E_L$, the instantiated ``effective" connectivity matrix is $\hat{E}_L$:
\begin{equation*} P(\hat{E}_L|E_L) = p^{K(\hat{E}_L)}(1-p)^{K(E_L-\hat{E}_L)}, \end{equation*}
where $K(M)$ is the number of nonzero elements in matrix $M$. Each element of the transition matrix in pattern space is then given by
\begin{equation*} P(\mathbf x_L=x | \mathbf x_{L-1} = \tilde x, E_L ) = \sum_{\hat{E}_L} P(\hat{E}_L|E_L) \cdot \mathbb{I} \left[ x = \Theta ( \hat{E}_L \tilde{x} - \theta ) \right]. \end{equation*}
where $\mathbb{I}$ denotes the indicator function. We will in two steps reduce the dimension of the system to condense pattern space into rate space. First, summing over $x$:
\begin{equation*} P(S_L=m | \mathbf x_{L-1} = \tilde{x},E_L) = \sum_{x} P(\mathbf x_L=x | \mathbf x_{L-1} = \tilde{x},E_L) \cdot \mathbb{I}\left[\sum_{i=1}^N x(i)=m\right]. \end{equation*}
Another sum gives the $(N+1)\times(N+1)$ transition matrix conditioned on the connectivity matrix $E_L$:
\begin{align*}
P(S_L=m|S_{L-1}=n,E_L) &= \sum_{\tilde{x}} P(S_L=m | \mathbf x_{L-1}=\tilde{x},S_{L-1},E_L) \cdot P(\mathbf x_{L-1} | S_{L-1},E_L)\\
&= \sum_{\tilde{x}} P(S_L=m | \mathbf x_{L-1}=\tilde{x},E_L) \cdot P(\mathbf x_{L-1} | S_{L-1}).
\end{align*}
Finally, averaging over every possible $E_L$ for the fixed $\gamma$, we obtain
\begin{equation*} P(S_L=m|S_{L-1}=n) = \sum_{E_L} P(S_L=m | S_{L-1}=n, E_L) \cdot P(E_L),\end{equation*}
which gives the elements of the mean-field transition matrix. Through these steps, the explicit derivation of the mean-field model from the original setup is demonstrated.

\subsection*{Validity of the mean-field Markov chain model}
\begin{figure}[h!]
\hfill
\begin{center}
\includegraphics[width = 5.8 in] {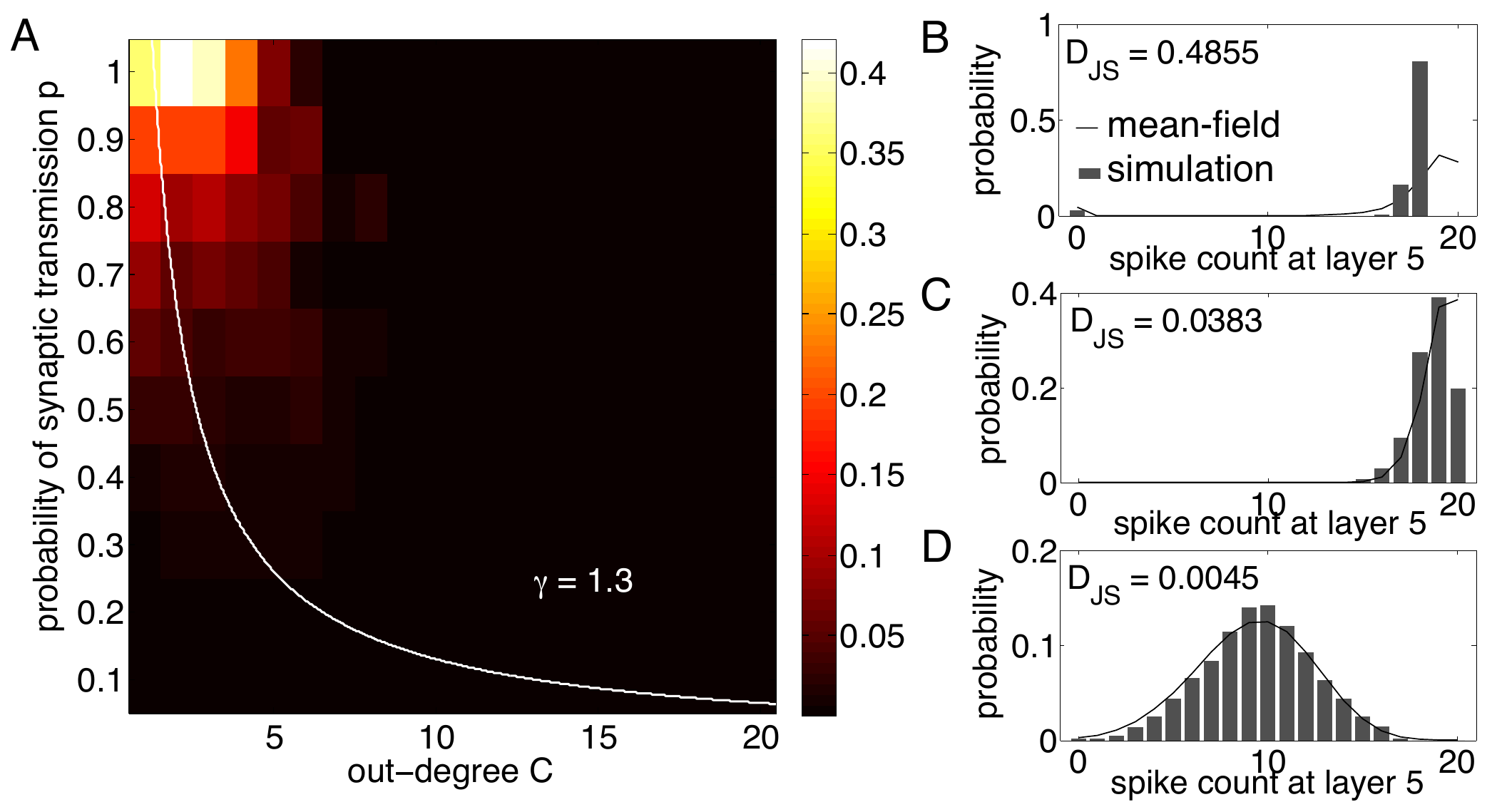}
\end{center}
\caption{Investigating the validity of the mean-field model. (A) Average JS divergence between the distribution after simulation through five layers and that predicted by the mean-field model for varying $C$, $p$. The mean-field model breaks down in the deterministic limit of small $C$ and high $p$. The white curve represents $\gamma \approx \gamma_{\text{eig}}\approx 1.3$. (B - D) Example spike count distributions from 1000 Monte Carlo simulations (grey bars) and their mean-field predictions (black line) for three orders of magnitude of the JS divergence. Parameters are (B) $C = 3$, $p = 1$, $S = 3$ for the worst fit, (C) $C = 6$, $p = 0.5$, $S = 9$ for the intermediate fit, and (D) $C = 5$, $p = 0.26$, $S = 11$ for the best fit.}
\label{mfcheck}
\end{figure}

In this section we investigate the validity of the mean-field Markov chain model.  Specifically, for a fixed network connectivity structure, we first estimate the true spike count distribution $P_5$ in response to input rate $S$ through Monte Carlo simulation. We then compare this to the distribution predicted by the mean-field Markov chain $P_5^{\text{MF}} = P_{\text{input}} A^4$ by computing the Jensen-Shannon divergence between these two distributions. Finally, we average the JS divergence over 100 instantiations of all possible input rates and over 20 random networks for that particular $C$ and $p$.

Overall, the mean-field distribution approximates the true spike count distribution quite accurately, as shown in Figure~\ref{mfcheck}A. The white curve overlain on the figure indicates the level set $\gamma \approx \gamma_{\text{eig}}$.  Note that agreement is perfect for fully connected networks.   The only major challenge to the accuracy of the mean-field approximation is in the deterministic limit of low $C$ and high $p$. Since $C$ is low there are few trials for the stochastic synapses, and the high $p$ additionally ensures that over repetitions of the same stimulus $S$ the activity follows a nearly deterministic trajectory, resulting in $P_5$ having a narrower distribution than the mean-field predicts. Example histograms are shown in Figure~\ref{mfcheck}BC to give an interpretation of values for the JS divergence. 

When repeated for $\theta = 7$ (data not shown), the mean-field model even better captured the true distributions, with a maximal JS divergence of 0.15 in the region of inaccuracy in the limit of $p \approx 1$ and $C \approx \gamma_{obs}$. As a final check, we also compared the {\it means} of the response distributions and found that, as expected, the averaged error was below machine epsilon (results not shown).

\subsection*{Analytical results for the eigenstructure of the mean-field transition matrix}
\begin{proposition} For any threshold $\theta \geq 1$ and connectivity $0<\gamma < N$, the transition matrix $A$ possesses a unique stationary state $\pi=v_{\emph{off}}$ such that $\pi A = \pi$. \end{proposition} 
\noindent {\bf Proof:} Let $\pi = (p_0, \dots, p_N )$. Then from direct matrix multiplication, the $m$th component of the vector $\pi A$ is
\begin{equation*} (\pi A)_m = \sum_{n=0}^N p_n {N \choose m} q_n^m (1-q_n)^{N-m}. \end{equation*}
The stationary state requires $p_m = (A p)_m$ for all $m$, i.e. 
\begin{equation*} p_m = \sum_{n=0}^N p_n {N \choose m} q_n^m (1-q_n)^{N-m} \end{equation*}
for all $m = 0, \dots, N$. In particular, when $m=0$, this becomes 
\begin{equation*} p_0 = p_0 + \sum_{n=1}^N p_n {N \choose 0} q_n^0 (1-q_n)^{N}. \end{equation*}
The summed term on the right hand side must be zero. However, note that each of the components of this sum is nonnegative, so they each must be zero, i.e., for each $n = 1, \dots, N$ either $p_n = 0$ or $(1-q_n)^N = 0$. We could have $(1-q_n)^N=0$ for a particular $n$ if $q_n=1$. However, $q_n$ can never be $1$ for sensible parameter values of $\theta > 0$ and $0 < \gamma < N$. Therefore, we must have $p_n = 0$ for all $n = 1, \dots, N$, and thus $p_0 = 1$.  The resulting stationary state is therefore unique and precisely equal to $v_{\text{off}}$. $\square$

\begin{proposition} Suppose the  eigenvectors of $A$ have limits as $\gamma \to N$. Then, $A$ has an eigenvalue $\lambda^* \to 1$ as $\gamma \to N$ with corresponding eigenvector $v^*$ that converges to a vector in the span of $v_{\emph{on}}$ and $v_{\emph{off}}$. \end{proposition}
\noindent {\bf Proof:} First consider 
\begin{equation*} q_n = 1 - \sum_{k=0}^{\theta-1} {n \choose k} \left( \frac{\gamma}{N}\right)^k \left(1-\frac{\gamma}{N} \right)^{n-k} \end{equation*}
as $\gamma \to N$. For $n\leq \theta$, $q_n =0$ by definition. For $n>\theta$, the sum on the right side of this equation approaches $0$ since $n>k$, so $q_n \to 1$. Below we summarize for various $m$ and $n$ the limit of $q_n^m (1-q_n)^{N-m}$ as $\gamma \to N$:
\begin{align*}
n>\theta: \hspace{.2 in} m=0: \hspace{.2 in} &q_n^0(1-q_n)^N \to 0 \\
0<m<N: \hspace{.2 in} &q_n^m(1-q_n)^{N-m} \to 0 \\
m=N: \hspace{.2 in}  &q_n^N(1-q_n)^0 \to 1 \\
n\leq \theta: \hspace{.2 in} m=0: \hspace{.2 in} &q_n^0(1-q_n)^N \to 1 \\
0<m<N: \hspace{.2 in} &q_n^m(1-q_n)^{N-m} \to 0 \\
m=N: \hspace{.2 in} &q_n^N(1-q_n)^0 \to 0.
\end{align*}
Now suppose $\lambda$ is an eigenvalue of $A$ with corresponding eigenvector $v$ for some $\gamma$. Then, $\lambda$ and $v$ satisfy
\begin{equation*} \sum_{n=0}^N v_n {N \choose m} q_n^m (1-q_n)^{N-m} = \lambda v_m\end{equation*}
 for all $m=0,\dots,N$. In particular, for $m=0$, we have:
\begin{equation*} \sum_{n=0}^{\theta} v_n {N \choose 0} q_n^0(1-q_n)^N + \sum_{n=\theta+1}^N v_n {N \choose 0} q_n^0(1-q_n)^N = \lambda v_0,\end{equation*}
which, taking $\gamma \to N$, reduces to the following:
\begin{equation*} \sum_{n=0}^{\theta} \tilde{v}_n = \tilde{\lambda} \tilde{v}_0,\end{equation*} 
all other terms having vanished.
Here, $\tilde{v}$ is the limit of $v$, which exists by assumption, and $\tilde{\lambda}$ is the limit of $\lambda$, which exists by the continuity of eigenvalues. For $m=N$, a similar expression is obtained:
\begin{equation*} \sum_{n=\theta+1}^N \tilde{v}_n = \tilde{\lambda} \tilde{v}_N. \end{equation*}
Finally, for $0<m<N$:
\begin{equation*} 0 = \tilde{\lambda} \tilde{v}_m \end{equation*}
This last equation reveals two possibilities: either $\tilde{\lambda} =0$ or $\tilde{v}_m=0$ for $0<m<N$. The latter case implies that $\lambda = 1$, thus the second largest eigenvalue of $A$ converges to $1$ with limiting eigenvector in the span of $v_{\text{on}}$ and $v_{\text{off}}$. All other eigenvalues converge to $0$. $\square$

\begin{proposition} Suppose $N_E - N_I > \theta$. Then, the (four-dimensional) transition matrix $A_{\text{in}}$ 
has a unique (two-dimensional) stationary state $\pi$ 
corresponding to no inhibitory and no excitatory neurons spiking at a layer. Moreover, the second largest eigenvalue converges to 1, and assuming the eigenvectors of $A_{\text{in}}$ have limits as $\gamma \to N = N_E+N_I$, then its corresponding eigenvector converges to the space spanned by the vector corresponding to all inhibitory and excitatory neurons firing, and the vector corresponding to all inhibitory and excitatory neurons being quiescent. \end{proposition}
\noindent {\bf Proof:} Because of the structure of $A_{\text{in}}$, this proposition follows similarly to those of the previous two propositions. $\square$

\subsection*{Another metric for rate propagation}
\begin{figure}[h!]
\hfill
\begin{center}
\includegraphics[width = 5.8 in] {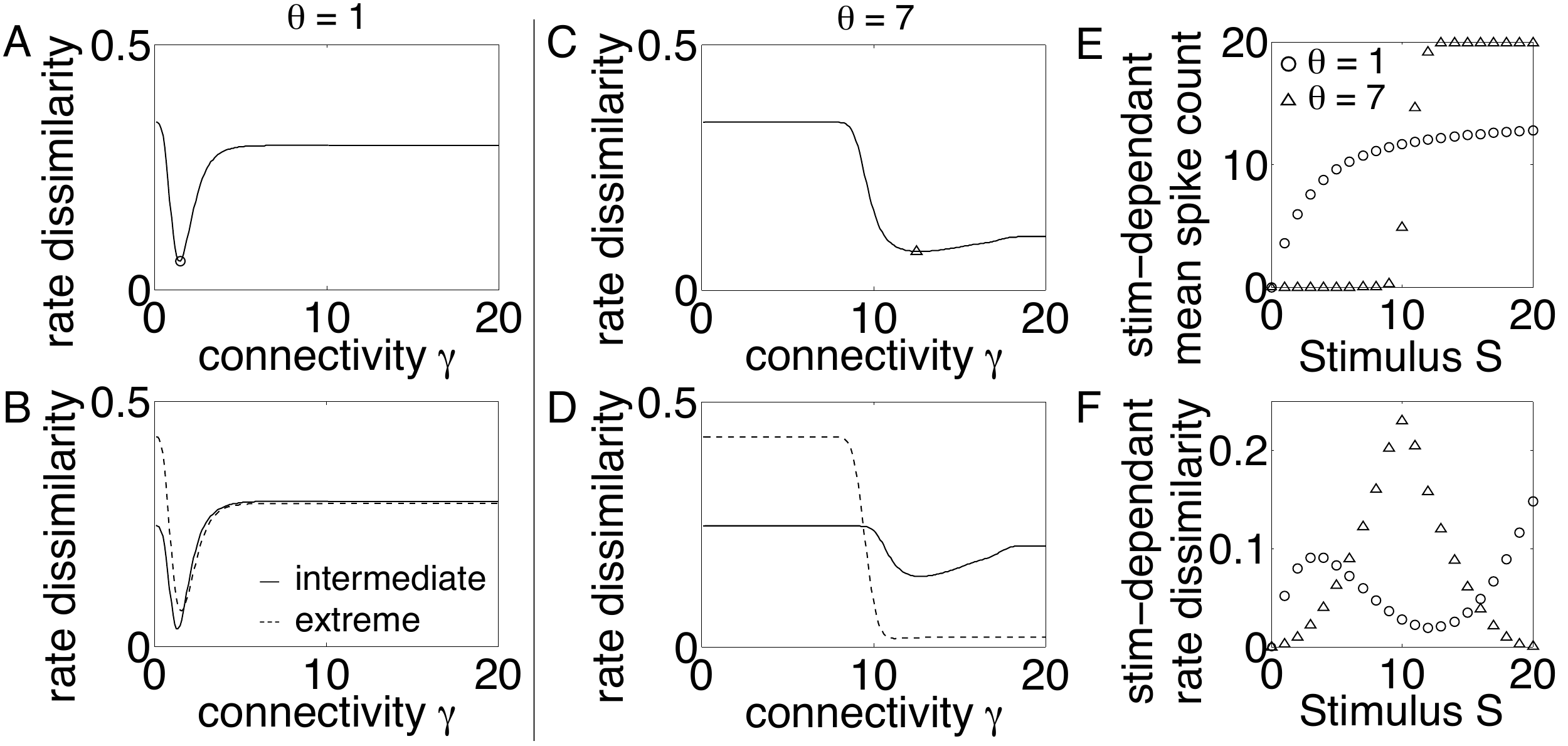} 
\end{center}
\caption{Rate dissimilarity for (A, B) $\theta = 1$, and (C, D) $\theta = 7$. (A, C) Rate dissimilarity plotted as a function of connectivity level $\gamma$. For (A) low-threshold networks, rate propagation is optimal at $\gamma_\text{obs}$. For (C) high-threshold networks, this is no longer the case. (B, D) Rate dissimilarity averaged over intermediate values ($S = 6,\dots,15$, solid line) and extreme rates ($ S = 0,\dots,5$ and $16,\dots,20$, dashed line). (E) Mean of spike count and (F) rate dissimilarity for high-threshold (triangles) and low-threshold (circles) networks plotted as a function of input stimulus. The networks shown in (E) and (F) are those that minimize the stimulus-averaged rate dissimilarity, as indicated by the markers in (A, C).}
\label{ratedissim}
\end{figure}

In addition to the measures described in the main text, we also considered the metric for rate propagation following \citet{van Rossum}. Define the \emph{rate dissimilarity} between the input rate $S/N$ and the rate at the $L$th layer $S_L/N$ via:
\begin{equation*} \text{RD}(\gamma,\theta) = \mathbb E_S \left[ \mathbb E_{\text{trials}} \left  [\left(S_L/N-S/N\right)^2 \middle | S \right ]\right]. \end{equation*}
There are two potential sources of poor performance according this quantification: (1) if the mean value of $S_L$ is far from $S$, or (2) if $S_L$ has large variance. As we see in Figure~\ref{ratedissim}A, when $\theta = 1$ the rate dissimilarity reaches its minimal value at critical connectivity $\gamma_{\text{eig}} \approx \gamma_{\text{obs}}$, suggesting that for low-threshold neurons, these networks are best able to propagate rates through the network. Outside of this intermediate connectivity range, the dissimilarity between input and output returns to high values. 

When threshold is raised, the dissimilarity curve changes shape and no longer has a sharp minimum at $\gamma_{\text{eig}}$ (Figure~\ref{ratedissim}C); instead, there is a robust minimum. Moreover, the minimal rate dissimilarity values for the low- and high-threshold networks are at comparable values. This may at first seem surprising, given that the high-threshold networks produce strong synchrony, and this should lead to large response variance. What is actually happening is an effect of both the increasing nullity of $A$ and averaging over all stimuli. In Figure~\ref{ratedissim}E the stimulus-dependent mean of the output at the 5th layer is plotted as a function of the stimulus for the networks that minimize average rate dissimilarity, indicated by the markers in Figure~\ref{ratedissim}AC, for both $\theta = 1$ (circles) and $\theta = 7$ (triangles). It is immediately clear that the low-threshold network better propagates intermediate rates as compared to the high-threshold network. By calculating the stimulus-dependent rate dissimilarity, rather than taking the uniform average, we see in Figure~\ref{ratedissim}F the difference between these two networks.  While high-threshold networks can propagate low and high rates better than low-threshold networks, only the latter can propagate intermediate rates.  This is because high-threshold networks produce bimodal responses at the connectivity value required to propagate rates. To make this point more apparent, in Figure~\ref{ratedissim}BC we have crudely separated the rate dissimilarity averaged over intermediate rates (solid lines) and extreme (either high or low) rates (dashed lines). This reveals that low-threshold networks perform better than high-threshold networks for intermediate rates. 

Faced with the subtlety of these results, in the main text we use the spike count JS divergence in order to unambiguously reveal network properties that support the propagation of asynchronous input distributions.


\begin{thebibliography}{100}
\providecommand{\natexlab}[1]{#1}
\expandafter\ifx\csname urlstyle\endcsname\relax
  \providecommand{\doi}[1]{doi:\discretionary{}{}{}#1}\else
  \providecommand{\doi}{doi:\discretionary{}{}{}\begingroup
  \urlstyle{rm}\Url}\fi
  
\bibitem[{Amari et~al.(2003)Amari, Nakahara, Wu, \& Sakai}]{Amari} 
Amari, S., Nakahara, H., Wu, S., and Sakai, Y. (2003).
\newblock Synchronous firing and higher-order interactions in neuron pool.
\newblock \emph{Neural Computation}, \emph{15(1)}, 127-142.

\bibitem[{Beggs \& Plenz(2003)Beggs, \& Plenz}]{Beggs}
Beggs, J., and Plenz, D. (2003).
\newblock Neuronal avalanches in neocortical circuits.
\newblock  \emph{The Journal of Neuroscience}, \emph{23(35)}, 11167-11177.

\bibitem[{Bertschinger \& Natschlager(2004)Bertschinger, \& Natschlager}]{Bertschinger} 
Bertschinger, N., and Natschlager, T. (2004). 
\newblock Real-time computation at the edge of chaos in recurrent neural networks. 
\newblock \emph{Neural Computation}, \emph{16(7)}, 1413-1436.

\bibitem[{Braitenberg \& Sch{\"u}z(1998)Braitenberg, \&  Sch{\"u}z}]{Braitenberg}
Braitenberg, V., and Sch{\"u}z, A. (1998).
\newblock \emph{Cortex: Statistics and Geometry of Neuronal Connectivity}. Berlin: Springer.

\bibitem[{Ganmor et~al.(2011)Ganmor, Segev, \& Schneidman}]{Ganmor}
Ganmor, E., Segev, R., and Schneidman, E. (2011).
\newblock Sparse low-order interaction network underlies a highly correlated and learnable population code.
\newblock \emph{Proceedings of the National Academy of Sciences USA}, \emph{108}, 9679--9684.

\bibitem[{Hahn et~al.(2010)Hahn, Petermann, Havenith, Yu, Plenz, Singer, \& Nikolic}]{Hahn} 
Hahn, G., Petermann, T., Havenith, M., Yu, S., Plenz, D., S
ger, W., and Nikolic, D. (2010). 
\newblock Neuronal avalanches in spontaneous activity in vivo. 
\newblock \emph{Journal of Neurophysiology}, \emph{104(6)}, 3312-3322.

\bibitem[{Haldeman \& Beggs(2005)Haldeman, \& Beggs}]{Haldeman}
Haldeman, C., and Beggs, J. (2005). 
\newblock Critical branching captures activity in living neural networks and maximizes the number of metastable states. 
\newblock \emph{Physical Review Letters}, \emph{94(5)}, 058101.

\bibitem[{Hennig et~al.(2009)Hennig, Adams, Willshaw, \& Sernagor}]{Hennig} 
Hennig, M. H., Adams, C., Willshaw, D., and Sernagor, E. (2009).
\newblock Early-stage waves in the retinal network emerge close to a critical state transition between local and global functional connectivity.
\newblock \emph{The Journal of Neuroscience}, \emph{29(4)}, 1077-1086.

\bibitem[{Jaynes(1957)Jaynes}]{Jaynes}
Jaynes, E.T. (1957).
\newblock Information theory and statistical mechanics.
\newblock \emph{The Physical Review}, \emph{106}, 620-630.

\bibitem[{Kumar et~al.(2010)Kumar, Rotter, \& Aertsen}]{Kumar} 
Kumar, A., Rotter, S., and Aertsen, A. (2010). 
\newblock Spiking activity propagation in neuronal networks: reconciling different perspectives on neural coding.
\newblock \emph{Nature Reviews Neuroscience}, \emph{11(9)}, 615-627.

\bibitem[{Kitzbichler et~al.(2009)Kitzbichler, Smith, Christensen, \& Bullmore}]{Kitzbichler}
Kitzbichler, M. G., Smith, M. L., Christensen, S. R., and Bullmore, E. (2009).
\newblock Broadband criticality of human brain network synchronization.
\newblock \emph{PLoS Computational Biology}, \emph{5(3)}, e1000314.

\bibitem[{Legenstein \& Maass(2007)Legenstein, \& Maass}]{Legenstein} 
Legenstein, R., and Maass, W. (2007).
\newblock What makes a dynamical system computationally powerful?
\newblock In S. Haykin, J. C. Principe, T. J. Sejnowski, and J. G. McWhirter (Eds.), \emph{New Directions in Statistical Signal Processing: From Systems to Brains} (pp. 127-154). Cambridge, MA: MIT Press.

\bibitem[{Litvak et~al.(2003)Litvak, Sompolinsky, Segev, \& Abeles}]{Litvak}
Litvak, V., Sompolinsky, H., Segev, I., and Abeles, M. (2003).
\newblock On the transmission of rate code in long feedforward networks with excitatory-inhibitory balance.
\newblock \emph{The Journal of Neuroscience}, \emph{23}, 3006-3015.

\bibitem[{Longtin(1993)Longtin}]{Longtin}
Longtin, A. (1993).
\newblock Stochastic resonance in neuron models.
\newblock \emph{The Journal of Statistical Physics}, \emph{70}, 309-327.

\bibitem[{Macke et~al.(2011)Macke, Opper, \& Bethge}]{Macke} 
Macke, J., Opper, M., and Bethge, M. (2011).
\newblock Common input explains higher-order correlations and entropy in a simple model of neural population activity.
\newblock \emph{Physical Review Letters}, \emph{106(20)}, 208102.

\bibitem[{Martignon et~al.(2000)Martignon, Deco, Laskey, Diamond, Freiwald, \& Vaadia}]{Martignon}
Martignon, L., Deco, G., Laskey, K., Diamond, M., Freiwald, W., and Vaadia, E. (2000).
\newblock Neural coding: higher-order temporal patterns in the neurostatistics of cell assemblies.
\newblock \emph{Neural Computation}, \emph{12(11)}, 2621-2653.

\bibitem[{McCulloch \& Pitts(1943)McCulloch, \& Pitts}]{McCulloch} 
McCulloch, W. S., and Pitts, W. H. (1943).
\newblock A logical calculus of the ideas immanent in nervous activity.
\newblock \emph{Bulletin of Mathematical Biophysics}, \emph{7}, 115Ð133.

\bibitem[{McDonnell \& Abbott(2009)McDonnell, \& Abbott}]{McDonnell} 
McDonnell, M., and Abbott, D. (2009). 
\newblock What is stochastic resonance? Definitions, misconceptions, debates, and its relevance to biology.
\newblock \emph{PLoS Computational Biology}, \emph{5(5)}, e1000348.

\bibitem[{Montani et~al.(2009)Montani, Ince, Senatore, Arabzadeh, Diamond, \& Panzeri}]{Montani}
Montani, F., Ince, R. A. A., Senatore, R., Arabzadeh, E., Diamond, M. E., and Panzeri, S. (2009).
\newblock The impact of high-order interactions on the rate of synchronous discharge and information transmission in somatosensory cortex.
\newblock \emph{Philosophical Transactions of the Royal Society A: Mathematical, Physical and Engineering Sciences}, \emph{367(1901)}, 3297--3310.

\bibitem[{Mora \& Bialek(2011)Mora, \& Bialek}]{Mora} 
Mora, T., and Bialek, W. (2011).
\newblock Are biological systems poised at criticality?
\newblock \emph{The Journal  of Statisitical Physics}, \emph{144(2)}, 268-302.

\bibitem[{Natschlager et~al.(2005)Natschlager, Bertschinger, \& Legenstein}]{Natschlager} 
Natschlager, T., Bertschinger, N., and Legenstein, R. (2005).
\newblock At the edge of chaos: real-time computations and self-organized criticality in recurrent neural networks.
\newblock In  L. Saul, Y. Weiss, and L. Bottou (Eds.), \emph{Advances in Neural Information Processing Systems 17} (pp. 142-152). Cambridge, MA: MIT Press.

\bibitem[{Nowotny \& Huerta(2003)Nowotny, \& Huerta}]{Nowotny} 
Nowotny, T., and Huerta, R. (2003). 
\newblock Explaining synchrony in feed-forward networks.
\newblock \emph{Biological Cybernetics}, \emph{89(4)}, 237-241.

\bibitem[{Petermann et~al.(2009)Petermann, Thiagarajan, Lebedev, Nicolelis, Chialvo, \& Plenz}]{Petermann} 
Petermann, T., Thiagarajan, T. C., Lebedev, M. A., Nicolelis, M. A., Chialvo, D. R., and Plenz, D. (2009).
\newblock Spontaneous cortical activity in awake monkeys composed of neuronal avalanches.
\newblock \emph{Proceedings of the National Academy of the Sciences}, \emph{106(37)}, 15921-15926.

\bibitem[{Reyes(2003)}]{Reyes} 
Reyes, A. (2003).
\newblock Synchrony-dependent propagation of firing rate in iteratively constructed networks in vitro.
\newblock \emph{Nature Neuroscience}, \emph{6(6)}, 593-599.

\bibitem[{Rosenbaum et~al.(2010)Rosenbaum, Trousdale, \& Josic}]{Rosenbaum}
Rosenbaum, R., Trousdale, J., and Josic, K. (2010).
\newblock Pooling and correlated neural activity.
\newblock \emph{Frontiers in Computational Neuroscience}, \emph{4(9)}, doi: 10.3389/fncom.2010.00009.

\bibitem[{van Rossum et~al.(2002)van Rossum, Turrigiano, \& Nelson}]{van Rossum} 
van Rossum, M., Turrigiano, G., and Nelson, S. (2002). 
\newblock Fast propagation of firing rates through layered networks of noisy neurons.
\newblock \emph{The Journal of Neuroscience}, \emph{22(5)}, 1956-1966.

\bibitem[{Schneidman et~al.(2006)Schneidman, Berry, Segev, \& Bialek}]{Schneidman}
Schneidman, E., Berry, M., Segev, R., and Bialek, W. (2006).
\newblock Weak pairwise correlations imply strongly correlated network states in a neural population.
\newblock \emph{Nature}, \emph{440(20)}, 1007-1012.


\bibitem[{Shlens et~al.(2006)Shlens, Field, Gauthier, Grivich, Petrusca, Sher, Litke, \& Chichilinsky}]{Shlens}
Shlens, J., Field, G.D., Gauthier, J.L., Grivich, M.I., Petrusca, D., Sher, A., Litke, A.M., and Chichilnisky, E.J. (2006).
\newblock The structure of multi-neuron firing patterns in primate retina.
\newblock \emph{The Journal of Neuroscience}, \emph{26(32)}, 8254-8266.

\bibitem[{Staude et~al.(2010).Staude, Rotter, \& Gr{\"u}n}]{Staude}
Staude, B., Rotter, S., and Gr{\"u}n, S. (2010).
\newblock CuBIC: cumulant based inference of higher-order correlations in massively parallel spike trains.
\newblock \emph{The Journal of Computational Neuroscience}, \emph{29}, 327Ð350.

\bibitem[{Trefethen \& Embree(2005)Trefethen, \& Embree}]{Trefethen}
Trefethen, L. N., and Embree, N. (2005).
\newblock \emph{Spectra and Pseudospectra: The Behavior of Nonnormal Matrices and Operators}. Princeton, NJ: Princeton University Press.

\bibitem[{Yu et~al.(2011)Yu, Yang, Nakahara, Santos, Nikolic, \& Plenz}]{Yu}
Yu, S., Yang, H., Nakahara, H., Santos, G., Nikolic, D., and Plenz, D. (2011).
\newblock Higher-order interactions characterized in cortical activity.
\newblock \emph{The Journal of Neuroscience}, \emph{31(48)}, 17514-17526.




\end{thebibliography}
\end{document}